\documentclass[prd,twocolumn,showpacs,preprintnumbers,amsmath,amssymb,nofootinbib]{revtex4}
\pdfoutput=1

\usepackage{amsmath}
\usepackage{graphicx}
\usepackage{dcolumn}
\usepackage{xspace}
\usepackage{color}
\usepackage{url}
\usepackage[utf8]{inputenc}
\usepackage{epstopdf}

\newcommand{\bea}{\begin{eqnarray}}
\newcommand{\eea}{\end{eqnarray}}
\newcommand{\nn}{\nonumber\\}

\newcommand{\eq}[1]{Eq.~\eqref{#1}}

\begin{document}
\preprint{PSI-PR-17-13}
\title{A model of vector leptoquarks in view of the $B$-physics anomalies}

\author{Lorenzo Calibbi}
\email{calibbi@itp.ac.cn}
\affiliation{CAS Key Laboratory of Theoretical Physics, Institute of Theoretical Physics, 
Chinese Academy of Sciences, Beijing 100190, China}
\author{Andreas Crivellin}
\email{andreas.crivellin@cern.ch}
\affiliation{Paul Scherrer Institut, CH--5232 Villigen PSI, Switzerland}
\author{Tianjun Li}
\email{tli@itp.ac.cn}
\affiliation{CAS Key Laboratory of Theoretical Physics, Institute of Theoretical Physics, Chinese Academy of Sciences, Beijing 100190, China}
\affiliation{School of Physical Sciences, University of Chinese Academy of Sciences, No.~19A Yuquan Road, Beijing 100049, China}

\begin{abstract}
Lepton number as a fourth color is an intriguing theoretical idea which is combined with a possible left-right symmetry within the famous Pati-Salam (PS) model. In the conventional PS model, a spontaneous breaking of the PS gauge group down to the SM one can only take place at very high scales (above the PeV scale) due to the stringent bounds from $K_L\to\mu e$ and $K\to\pi\mu e$ induced by the resulting vector leptoquarks. In this paper, we show that these constraints can be avoided once additional vector-like fermions are introduced and thus a breaking at the TeV scale is possible. We consider the flavor phenomenology of this model in the context of the intriguing hints for new physics in semileptonic $B$ decays. The necessary violation of lepton flavor universality is induced by mixing SM and vector-like fermions. Concerning $R(D)$ and $R(D^*)$ we find that sizable effects are possible while respecting the bounds from other flavor observables but predicting a large enhancement of $B_s\to\tau^+\tau^-$. Furthermore, also in $b\to s\ell^+\ell^-$ transitions the observed deviations from the SM predictions (including $R(K)$ and $R(K^*)$) can be explained with natural values for the free parameters of the model without any fine-tuning, predicting sizable decay rates for $b\to s\tau\mu$. Finally, the anomaly in anomalous magnetic moment of the muon can be accounted for by a loop-contribution involving the vector leptoquark and vector-like leptons.
\end{abstract}
\pacs{13.25.Hw,14.80.Sv}

\maketitle

\section{Introduction}
\label{intro}

Until now, the Large Hadron Collider at CERN did not directly observe any particles beyond the ones present in the Standard Model (SM) of particle physics. However, we have accumulated intriguing hints for lepton flavor universality (LFU) violation in semi-leptonic $B$ decays within recent years. Most prominently, there exist deviations from the SM predictions in $b\to s\mu^+\mu^-$ above the $5\,\sigma$ level~\cite{Capdevila:2017bsm}\footnote{Including only $R(K)$ and $R(K^*)$ in the fit the significance is at the 4$\,\sigma$ level~\cite{Altmannshofer:2017yso,DAmico:2017mtc,Geng:2017svp,Ciuchini:2017mik,Hiller:2017bzc,Alok:2017sui,Hurth:2017hxg}.} and the combination of the ratios $R(D)$ and $R(D^*)$ differs by $4.1\,\sigma$ from its SM prediction~\cite{Amhis:2016xyh}. While $R(D)$ and $R(D^*)$ measure LFU violation in the charged current $B$ decays of the type $b\to c \ell \nu$, the fit to the $b\to s\ell^+\ell^-$ data also suggests the violation of LFU due to the measurement of $R(K)$~\cite{Aaij:2014ora} and $R(K^*)$~\cite{Aaij:2017vbb}. This implies a possible connection between these two classes of anomalies and motivates to search for a simultaneous explanation~\cite{Bhattacharya:2014wla,Calibbi:2015kma,Fajfer:2015ycq,Greljo:2015mma,Barbieri:2015yvd,Bauer:2015knc,Boucenna:2016qad,Das:2016vkr,Becirevic:2016yqi,Sahoo:2016pet,Bhattacharya:2016mcc,Barbieri:2016las,Alok:2017jaf,Crivellin:2017zlb,Chen:2017hir,Dorsner:2017ufx,Buttazzo:2017ixm}. 

The vector leptoquark $SU(2)$ singlet with hypercharge $-4/3$ is a natural candidate for a simultaneous explanation of $R(D)$ and $R(D^*)$ together with $b\to s\ell^+\ell^-$ data. First of all, it automatically fulfills the requirement that down-quarks do not couple at tree level to neutrinos and therefore avoids the stringent bounds from $B\to K^{(*)}\nu\nu$. This allows for large flavor violating couplings to quarks such that one can get a sizable effect in $R(D)$ and $R(D^*)$ with TeV scale masses such that the bounds from direct searches~\cite{Faroughy:2016osc} as well as from electroweak (EW) precision data~\cite{Feruglio:2016gvd} can be avoided~\cite{Crivellin:2017zlb}. In addition, unlike models with charged Higgses~\cite{Crivellin:2012ye,Tanaka:2012nw,Celis:2012dk,Crivellin:2013wna,Crivellin:2015hha,Chen:2017eby}, the vector leptoquark leaves the $q^2$ distribution in $R(D^{(*)})$ invariant (which is in good agreement with data~\cite{Freytsis:2015qca,Celis:2016azn,Ivanov:2017mrj}) and does not lead to an huge enhancement of $B_c\to \tau\nu$ which is incompatible with experiments~\cite{Li:2016vvp,Celis:2016azn,Alonso:2016oyd,Akeroyd:2017mhr}. Finally, it gives a $C_9=-C_{10}$-like effect in $b\to s\ell^+\ell^-$ transistions and therefore gives a good fit to data. However, a compelling renormalizable model giving rise to this leptoquark is still missing.

Interestingly, the vector leptoquark singlet with hypercharge $-4/3$ is contained within the theoretically very appealing PS model as a $SU(4)$ gauge boson. However, in the conventional model, the bounds on the symmetry breaking scale from $K_L\to\mu e$ and $K\to\pi\mu e$ are so strong (at the PeV scale)~\cite{Hung:1981pd,Valencia:1994cj} that any other observable effects in flavor physics are ruled out from the outset. Therefore, it must be extended if one aims at a realization at the TeV scale. In this article, we will construct a model based on the Pati-Salam gauge group in which the bounds from $K_L\to\mu e$ and $K\to\pi\mu e$ can be avoided. Furthermore, another crucial feature of the PS leptoquarks is that it allows for a low-energy realization since that it does not lead to proton decay at any loop level.

For this purpose, we add to the original PS model three pairs of fermions in the fundamental representation of $SU(4)$ with vector-like mass terms. These fermions can be considered as heavy vector-like generations. The mixing between them and the light SM particles is in general flavor dependent. Therefore, the model can have interesting effects in flavor physics, in particular, it could explain the hints for new physics in $b\to s\ell^+\ell^-$, $R(D^{(*)})$ and also the measurement of the anomalous magnetic moment (AMM) of the muon.

\section{The Model}
\label{sec:model}
Our starting point is the PS model~\cite{Pati:1974yy} with the gauge group $SU(4)\times SU(2)_L\times SU(2)_R$. Thus, left-handed fermions are $SU(2)_L$ doublets and right-handed fermions form $SU(2)_R$ doublets. This necessarily leads to the introduction of three right-handed neutrinos. In our model, we extend the fermion content of the original model having now 6 fermion fields $X^{L,R}_i,\,Y^{L,R}_i,\,Z^{L,R}_i$ as well as (at least) two more Higgs field $\Sigma^{1,2}$. These fields transform under the PS gauge group and one additional Peccei-Quinn-like $U(1)$ group as shown in Table \ref{fields}.
\begin{table}[h!]
{ \normalsize
\begin{tabular}{c|cccc}
& $SU\left( 4 \right)$ & $SU{{\left( 2 \right)}_L}$ & $SU{{\left( 2 \right)}_R}$ & $U{{\left( 1 \right)}_{PQ}}$\\
\hline
${X^{L}_i}$& 4&2&1&0\\
${Y^{L}_i}$& 4&2&1&-1\\
${Y^{R}_i}$& 4&2&1&1\\
${X^{R}_i}$& 4&1&2&0\\
${Z^{R}_i}$& 4&1&2&-1\\
${Z^{L}_i}$& 4&1&2&1\\
$\Sigma^X_1$ & ${\bar 4 \otimes 4}$ &1&1&{ - 1}\\
$\Sigma^X_2$ & ${\bar 4 \otimes 4}$ &1&1&{ - 1}\\
$\Sigma^Y_1$ & ${\bar 4 \otimes 4}$ &1&1&{ - 2}\\
$\Sigma^Y_2$ & ${\bar 4 \otimes 4}$ &1&1&{ - 2}
\end{tabular}
}
\caption{Field content of the model. Alternatively, one could use instead of $\Sigma^{X,Y}_{1,2}$ two fields $\Sigma^X_{ij}$, which transforms as $\bar 3 \otimes 3$ under a possible flavor symmetry.}
\label{fields}
\end{table}

\noindent Here the superscripts $L$ and $R$ label the chirality of the fields and $i=1,2,3$ is a flavor index. In the absence of the fields $Y$ and $Z$, the fields $X$ would be chiral fermions resembling the SM fermions.

In the following we will not explicitly specify the EW symmetry breaking sector whose Higgs fields are therefore not included in Table~\ref{fields}. However, we know that due to the decoupling theorem, the symmetry breaking sector must reduce, in the limit of heavy additional Higgses, to one light $SU(2)$ doublet with vev $v$ giving rise the chiral fermion and weak gauge boson masses. A possible completion of the above-sketched model, including the EW-breaking sector, will be given in Section \ref{sec:EWSB}. In our phenomenological discussion, we are not considering the implications of the extended Higgs sector, but rather we only include the pseudo-Goldstone bosons by working in unitary gauge. This approach is model independent in the sense that including additional physical Higgses would imply focusing on a specific UV realization of the model.

\subsection{Fermion masses}

Let us consider for simplicity only the $SU(2)_L$ doublet fermions ($X^{L}_i,\,Y_i^{L,R}$). The corresponding results for the $SU(2)_R$ follow in a straightforward way and they are not necessary for explaining the flavor anomalies as we will see later. Therefore, we can write down the following mass terms after the new scalar fields $\Sigma^{X,Y}_{1,2}$ acquire their vevs $v_{\Sigma^{X,Y}_{1,2}}$
\begin{align}
 \label{Lmass}
-{\cal L} ~\supset~&   {v_{\Sigma^X_1}^{ab} }x_{ij} \bar X_i^{aL}Y_j^{bR} + {v_{\Sigma^Y_1}^{ab} }y_{ij} \bar Y_i^{aL}Y_j^{bR}+ \\
& {v_{\Sigma^X_2}^{ab} }x^\prime_{ij} \bar X_i^{aL}Y_j^{bR} + {v_{\Sigma^Y_2}^{ab} }y^\prime_{ij} \bar Y_i^{aL}Y_j^{bR}+ h.c. \nonumber 
\end{align}
Here $a$ and $b$ are $SU(4)$ indices, and we denoted the Yukawa-like couplings by $x^{(\prime)}_{ij}$ and $y^{(\prime)}_{ij}$. Note that our assignment for the PQ charges was choosen in such a way that it avoids bare mass terms for the fermions before PS symmetry breaking. Therefore, the masses of the vector-like fermions are, for perturbative couplings, at most of the order of the $SU(4)$ breaking scale, which we assume to be around the TeV scale. After $\Sigma^{X,Y}_{1,2}$ acquire their vevs $SU(4)$ is broken down to $SU(3)_c\times U(1)_{B-L}$ and quarks and leptons become distinguishable. Decomposing the $SU(4)$ multiplets as
\begin{equation}
\label{LH-embedding}
{Y_R} = {\left( {\begin{array}{*{20}{c}}
{{Q'_R}}\\
{{L'_R}}
\end{array}} \right)_i},\;\;{Y_L} = {\left( {\begin{array}{*{20}{c}}
{{Q_L}}\\
{{\ell _L}}
\end{array}} \right)_i},\;\;{X_L} = {\left( {\begin{array}{*{20}{c}}
{{q_L}}\\
{{L_L}}
\end{array}} \right)_i}
\end{equation}
we see that $Q$ and $q$ are $SU(3)_c$ triplets corresponding to quarks, while $\ell$ and $L$ are $SU(3)_c$ singlets and thus correspond to leptons. Expanding \eq{Lmass} into components we find 
\begin{equation}
{\cal L} \supset - \left( {m_{ij}^Q{{\bar q}_{iL}} + M_{ij}^Q{{\bar Q}_{iL}}} \right){Q'_{jR}} - \left( {M_{ij}^L{{\bar L}_{iL}} + m_{ij}^L{{\bar \ell }_{iL}}} \right){L'_{jR}}\,,
\end{equation}
with 
\begin{equation}
\begin{array}{l}
m_{ij}^Q = {v_{\Sigma^X_1}^{11} }x_{ij}+{v_{\Sigma^X_2}^{11} }x^\prime_{ij}\,,\;\;m_{ij}^L = {v_{\Sigma^Y_1}^{22} }y_{ij}+{v_{\Sigma^Y_2}^{22} }y^\prime_{ij}\\
M_{ij}^L = {v_{\Sigma^X_1}^{22} }x_{ij}+{v_{\Sigma^X_2}^{22} }x^\prime_{ij}\,,\;\;M_{ij}^Q = {v_{\Sigma^Y_1}^{11} }y_{ij}+{v_{\Sigma^Y_2}^{11} }y^\prime_{ij}
\end{array}\,
\end{equation}
Here the superscript $11$ corresponds to a $3\times 3$ unit matrix in color space while $22$ represents only a single number. Here $v_{\Sigma^{X,Y}_{1,2}}^{12}=v_{\Sigma^{X,Y}_{1,2}}^{21}=0$ such that $SU(3)_c$ remains unbroken. We further assume $v_{\Sigma^{X,Y}_1}^{11}  \gg v_{\Sigma^{X,Y}_2}^{11}$ and $v_{\Sigma^{X,Y}_1}^{22}  \ll v_{\Sigma^{X,Y}_2}^{22}$, such that the mass terms
are dominantly given by
\begin{equation}
\begin{array}{l}
m_{ij}^Q \simeq {v_{\Sigma^X_1}^{11} }x_{ij}\,,\;\;m_{ij}^L \simeq {v_{\Sigma^Y_2}^{22} }y^\prime_{ij}\\
M_{ij}^L \simeq {v_{\Sigma^X_2}^{22} }x^\prime_{ij}\,,\;\;M_{ij}^Q \simeq {v_{\Sigma^Y_1}^{11} }y_{ij}
\end{array}\,
\end{equation}
Therefore, $M^{L,Q}_{ij}$ are the vector-like mass terms while $m^{L,Q}_{ij}$ provides the mixing of the vector-like fermions with the light (SM) ones. The study of the corresponding scalar potential is not trivial and requires future studies. 

Without loss of generality, one can choose $M^Q$ and $M^L$ to be diagonal in flavor space. In addition, we assume that $m^{Q,L}$ is diagonal in the same basis and for simplicity (without affecting the final results) that $M^{Q,L}$ is proportional to the unit matrix:
\begin{equation}
\label{eq:textures}
\begin{array}{l}
M_{ij}^{Q,L} = {M^{Q,L}}{\delta _{ij}}\,,\\
m_{ij}^{Q,L} = {\left( {\begin{array}{*{20}{c}}
{m_1^{Q,L}}&0&0\\
0&{m_2^{Q,L}}&0\\
0&0&{m_3^{Q,L}}
\end{array}} \right)_{ij}}\,.
\end{array}
\end{equation}
While the structure above is certainly not generic, it can be the consequence of an underlying flavor symmetry. In fact, if $Q_L$ and $Q_R'$ ($Q_L$ and $L_R'$) are triplets of $SU(3)$, 
$M^Q$ ($M^L$) does not break the symmetry, thus being proportional to the unit matrix. If on the contrary, $q_L$ and $\ell_L$ are anti-triplets, $m^{Q,L}$  are generated by $SU(3)$-breaking terms $\phi \phi /\Lambda $, where $\phi$ collectively denote the $SU(3)$-triplet scalar fields (`flavons') whose vevs break $SU(3)$ and $\Lambda$ a cutoff scale. The texture of $m^{Q,L}$ then follows from the flavor directions of the flavons' vevs. As an example, one can introduce two flavons, $\phi_3$ and $\phi_2$, with $\langle \phi_3 \rangle = (0,~0,~v_3)$, $\langle \phi_2 \rangle = (0,~v_2,~0)$. Distinguishing these two fields by an additional parity, the texture in Eq.~(\ref{eq:textures}) is obtained with 
$m_3^{Q,L}\sim v_3^2/\Lambda$, $m_2^{Q,L}\sim v_2^2/\Lambda$, $m_1^{Q,L}\sim0$. Therefore, the mixing with electrons is absent, and thus the vector LQs will not couple to them.\footnote{The absence of couplings to the electron at tree-level could alternatively be assured by an abelian flavour symmetry under which all fermions are equally charged except the electron. Furthermore, even though electron couplings will be generated at the loop level~\cite{Crivellin:2018yvo}, the absence of $\mu e$ couplings is RGE invariant and therefore no effect in $\mu\to e\gamma$~\cite{Crivellin:2017dsk} or $K\to \mu e$ is generated.}

Given the structure in Eq.~(\ref{eq:textures}), the mass matrices for quarks and leptons decompose each into three (one for each generation) rank one matrices diagonalized by the rotations
\begin{equation}
\begin{array}{l}
\left( {\begin{array}{*{20}{c}}
{{q_{iL}}}\\
{{Q_{iL}}}
\end{array}} \right) \to \left( {\begin{array}{*{20}{c}}
{{c_{iQ}}}&{ - {s_{iQ}}}\\
{{s_{iQ}}}&{{c_{iQ}}}
\end{array}} \right)\left( {\begin{array}{*{20}{c}}
{{q_{iL}}}\\
{{Q_{iL}}}
\end{array}} \right)\\
\left( {\begin{array}{*{20}{c}}
{{\ell _{iL}}}\\
{{L_{iL}}}
\end{array}} \right) \to \left( {\begin{array}{*{20}{c}}
{{c_{iL}}}&{ - {s_{iL}}}\\
{{s_{iL}}}&{{c_{iL}}}
\end{array}} \right)\left( {\begin{array}{*{20}{c}}
{{\ell _{iL}}}\\
{{L_{iL}}}
\end{array}} \right)
\end{array}\,.\label{rotations1}
\end{equation}
\begin{figure*}[t]
\begin{center}
\includegraphics[width=0.65\textwidth]{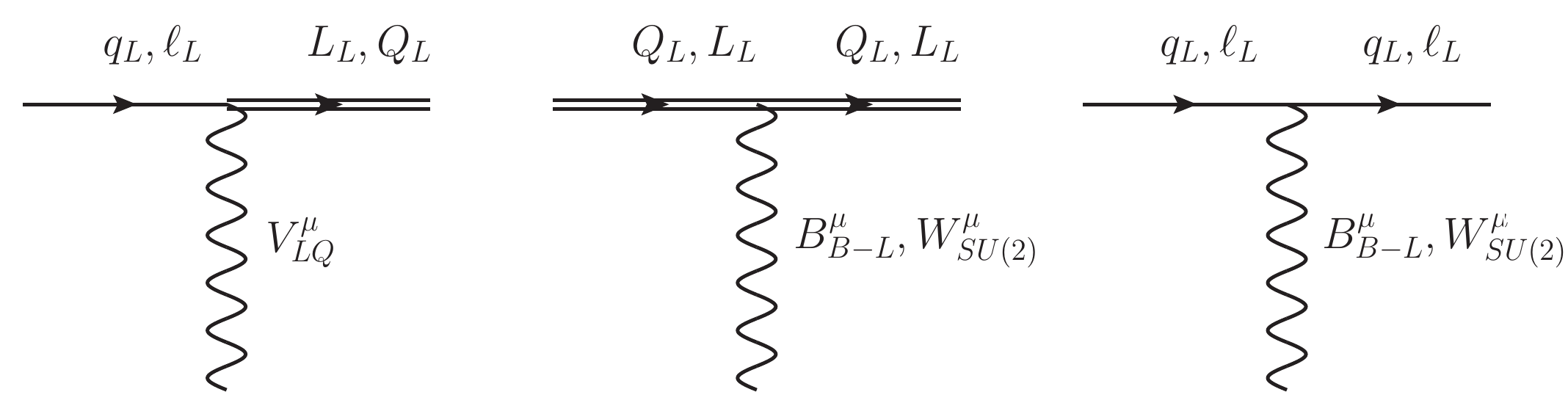}
\end{center}
\caption{Couplings of the gauge bosons to the heavy vector-like fermions ($L,Q$) and light SM-like fermions ($\ell,q$). After mixing among SM-like and vector-like fermions, the couplings to the leptoquark result become flavor non-universal, whereas the couplings to the other gauge bosons (in particular the one associated to $B-L$) remain flavor diagonal.}
\label{fig:diag}
\end{figure*}

As stated above, we do not explicitly specify the UV completion of the Higgs sector responsible for the EW symmetry breaking but rather use the decoupling theorem asserting that there is one light $SU(2)_L$ doublet with vev $v$ giving rise to the chiral fermion and weak gauge boson masses. We can now write down the usual Yukawa couplings and diagonalize the resulting $3\times 3$ matrices using biunitary transformations
\begin{equation}
{q_{iL}} \to U_{ij}^{qL}{q_{jL}},\;\;{\ell _{iL}} \to U_{ij}^{\ell L}{\ell _{jL}}\,,
\end{equation}
with $q=u,d$ and the corresponding expression for right-handed fields. For our final results, only the misalignment between left-handed quark and leptons 
\begin{equation}
U_{fi}^{q\ell L} = U_{jf}^{qL*}U_{ji}^{\ell L}\,,\label{rotations2}
\end{equation}
as well as the CKM matrix $V_{fi}^{\rm CKM} = U_{jf}^{uL*}U_{ji}^{dL}$ are important. Note that in the following, we work in the down basis, i.e. CKM rotations are only present once left-handed up-quarks are involved. We neglect Higgs couplings involving  chiral and vector-like fermions in our phenomenological analysis. 

In analogy to the $SU(2)_L$ sector, we embedded the fermions charged under $SU(2)_R$ in the following representations:
\begin{equation}
{Z_L} = {\left( {\begin{array}{*{20}{c}}
{{Q'_L}}\\
{{L'_L}}
\end{array}} \right)_i},\;\;{Z_R} = {\left( {\begin{array}{*{20}{c}}
{{Q_R}}\\
{{\ell _R}}
\end{array}} \right)_i},\;\;{X_R} = {\left( {\begin{array}{*{20}{c}}
{{q_R}}\\
{{L_R}}
\end{array}} \right)_i}
\label{RH-embedding}
\end{equation}
and the above discussion about masses and mixing can be replicated for the RH fermions of the SM.

\subsection{Couplings of fermions to gauge bosons}

After breaking of the $SU(4)$ symmetry, its 15 generators correspond to $8$ massless gluons, 6 leptoquarks ($V^\mu+\bar V^\mu$), and one $B-L$ gauge boson ($Z^{\prime\,\mu}$). As we can see from Fig.~\ref{fig:diag}, after mixing of $q$ ($\ell$) with $Q$ ($L$) the couplings of the $B-L$ gauge boson remains flavor universal with strength $\sqrt{{3}/{8}}\, g_s/3$ for  quarks, $\sqrt{{3}/{8}}\, g_s$ for leptons. The flavour universality of the $Z^\prime$ coupling is the result of the unitarity of the mixing matrices. Therefore, we did not need to assume any alignment between the vector-like mass terms and the Yukawa couplings to obtain this desirable feature.

Since we do not completely specify the Higgs sector (a possible realization is given in Section \ref{sec:EWSB}), we take the masses of the $B-L$ gauge boson and the leptoqaurks as free parameters. However, the masses should be of the same order, but due to the strong constraints from $Z^\prime$ searches,\footnote{With $g_{B-L}=\sqrt{{3}/{8}}\, g_s \approx 0.6$ at the TeV scale, we find that a recent ATLAS search for $Z' \to \ell^+ \ell^-$ \cite{Aaboud:2017buh} gives (in the narrow width approximation) $M_{Z'} \gtrsim 5$ TeV if the $Z'$ only decays into SM fermions. This is reduced to $M_{Z'} \gtrsim 4.6$ TeV in the more realistic case in which all decay channels of our $B-L$ $Z'$ into the vector-like fermions are open via a decrease of the $Z' \to \ell^+ \ell^-$ branching ratio by a factor $1/3$ and an increase of the total width up to $\approx 0.18\times M_{Z'}$.}
 we assume that the $B-L$ gauge boson is heavier than the leptoquarks. 
Such a mass splitting can be achieved if the PS symmetry breaking is due to the vev of a scalar in the symmetric representation of $SU(4)$. In fact, if the dominant contribution to the gauge boson masses is given by the field $\Phi$ of Table~\ref{fields-PS}, which is in the $(10,~1,~3)$ representation of the PS group, one obtains $M_{LQ}^2 = g_4^2 v_\Phi^2$ and $M_{Z'}^2 =3 g_4^2 v_\Phi^2 +2 g_R^2 v_\Phi^2$, which gives  $M_{Z'} \approx 2\times M_{LQ}$.
Furthermore, the limit on the $Z^\prime$ mass can be significantly weakened by introducing additional fermions which are charged under $B-L$ only, such as extra sterile neutrinos. This does not only decrease the branching ratio to muons and electrons (the modes which give the strongest bound  \cite{Aaboud:2017buh}) but also increases the total width, making the detection more difficult \cite{Faroughy:2016osc}. Finally, as we will see later, if we only aimed at a smaller effect in $b\to c\tau\nu$ processes rather than accounting for the central value, the gauge bosons (including the $Z^\prime$) can be heavier.

Let us now consider the couplings of the vector-leptoquark $V^\mu$. Here, the rotations in Eq.~(\ref{rotations1}) induced by the mixing between vector-like and SM fermions do not drop out, as it is apparent from Fig.~\ref{fig:diag}. In addition, after EW symmetry breaking, the misalignment between the rotations needed to diagonalize the light quark and lepton mass matrices $U^{q\ell L}_{fi}$, cf.~Eq.~(\ref{rotations2}), enters in the coupling of $V^\mu$ with the SM fermion doublets:
\begin{widetext}
\begin{equation}
{\cal  L} \supset -\frac{g_s}{\sqrt 2}U^{q\ell L}_{fi}{\left( {\begin{array}{*{20}{c}}
{\bar q_f^L}\\
{\bar Q_f^L}
\end{array}} \right)_a}{\gamma ^\mu }{P_L}{\left( {\begin{array}{*{20}{c}}
{\ell _i^L}\\
{L_i^L}
\end{array}} \right)_b}{\left( {\begin{array}{*{20}{c}}
{{c^Q_{{i}}}{s^L_{{i}}} + {c^L_{{i}}}{s^Q_{{i}}}}&{{c^L_{{i}}}{c^Q_{{i}}} - {s^L_{{i}}}{s^Q_{{i}}}}\\
{{c^L_{{i}}}{c^Q_{{i}}} - {s^L_{{i}}}{s^Q_{{i}}}}&{ - {c^Q_{{i}}}{s^L_{{i}}} - {c^L_{{i}}}{s^Q_{{i}}}}
\end{array}} \right)_{ab}}V_\mu ^{} +h.c.
\end{equation}
Considering for illustration only the second and third generations, which are of interest for our phenomenological study, we find
\begin{equation}
{\cal  L} \supset {\kappa _{ij}}\,\bar q_i^L{\gamma ^\mu }{P_L}\ell _j^L V_\mu ^{}+h.c.~~{\rm{with}}~~{\kappa _{ij}} = \frac{{ - {g_s}}}{{\sqrt 2 }}{\left( {\begin{array}{*{20}{c}}
{{c^Q_{{1}}}{s^L_{{1}}} + {c^L_{{1}}}{s^Q_{{1}}}}&0&0\\
0&{\left( {{c^Q_{{2}}}{s^L_{{2}}} + {c^L_{{2}}}{s^Q_{{2}}}} \right){c^{{q\ell}}_{23}}}&{ - {s{^{q\ell}_{23}}}\left( {{c^Q_{{2}}}{s^L_{{2}}} + {c^L_{{2}}}{s^Q_{{2}}}} \right)}\\
0&{\left( {{c^Q_{{2}}}{s^L_{{2}}} + {c^L_{{2}}}{s^Q_{{2}}}} \right){s^{ {q\ell}}_{23}}}&{{c^{ {q\ell}}_{23}}\left( {{c^Q_{{3}}}{s^L_{{3}}} + {c^L_{{3}}}{s^Q_{{3}}}} \right)}
\end{array}} \right)_{ij}}\,.
\end{equation}
\end{widetext}
Here $s^{{q\ell}}_{23}$, $c^{{q\ell}}_{23}$ are the rotations induced by the misalignment between the SM Yukawa couplings of quarks and leptons, encoded in~Eq.~(\ref{rotations2}). Recall that we assumed that the first generation quarks (in the interaction basis) and leptons do not mix with their SM partners. In this way, effects in $b\to d\ell^+\ell^-$ or Kaon decays~\cite{Crivellin:2016vjc} are suppressed and $\mu e$ lepton flavour violation is absent. This ensures that our model is consistent with the bounds from $K_L\to\mu e$ and $K\to\pi\mu e$ for TeV scale masses.
The suppression of the couplings of the vector leptoquark to first generation fermions can be accounted for by a flavor structure of the mixing mass terms $m^{Q,L}$ resembling the strong hierarchy of the SM Yukawas, which can be enforced by an underlying flavor symmetry, as in the $SU(3)$ example discussed below Eq.~(\ref{eq:textures}). In such a case, mixing with the first generation can be strongly suppressed by simply not introducing a flavon with a vev in the flavor direction $i=1$.

Similarly, couplings of $V^\mu$ to right-handed leptons and quarks might arise as an effect of the mixing with the vector-like fermions in the $SU(2)_R$ sector, i.e.~the field embedding in Eq.~(\ref{RH-embedding}). Such couplings should be small (but not necessarily zero) due to the observed patterns in $R(D^{(*)})$ and $b\to s\mu^+\mu^-$ transitions. In our setup, this can be easily achieved by a mild suppression of the SM-like/vector-like fermion mixing in the RH sector.
\section{Observables}
\subsection{$R(D)$ and $R(D^*)$}
We define the effective Hamiltonian for $b\to c\ell\nu$ transitions as
\begin{equation}
{H_{{\rm{eff}}}^{\ell_f\nu_i}} = \frac{{4{G_F}}}{{\sqrt 2 }}{V_{cb}}C_L^{fi}\left[ {\bar c{\gamma ^\mu }{P_L}b} \right]\left[ {{{\bar \ell }_f}{\gamma _\mu }{P_L}{\nu _i}} \right]\,,
\end{equation}
where in the SM $C_L^{fi}=\delta_{fi}$ and the contribution of our vector leptoquark is given by
\begin{equation}
{C_L} = \frac{{\sqrt 2 }}{{4{G_F}{V_{cb}}}}\frac{{\kappa _{33}^*V_{2j}^{}{\kappa _{j3}}}}{{{M^2}}}\,,
\end{equation}
leading to 
\begin{equation}
R(D^{(*)})/R(D^{(*)})_{\rm SM}=|1+C_L|^2\,,
\end{equation}
where we neglected contributions with muon or electron neutrinos. This has to be compared to the experimental measurements of
$R{\left( {{D^*}} \right)_{{\rm{EXP}}}} = 0.304 \pm 0.013\pm 0.007$ and $R{\left( D \right)_{{\rm{EXP}}}} = 0.407 \pm 0.039 \pm 0.024$, and the corresponding SM predictions, $R{\left( {{D^*}} \right)_{{\rm{SM}}}} = 0.252 \pm 0.003$ and $R{\left( D \right)_{{\rm{SM}}}} = 0.300 \pm 0.008$~\cite{Fajfer:2012vx,Na:2015kha}.

\subsection{$b\to s\ell^+\ell^-$ transitions}

Using the effective Hamiltonian
\begin{align}
H_{\rm eff}^{\ell_f\ell_i}&=- \dfrac{ 4 G_F }{\sqrt 2}V_{tb}V_{ts}^{*} \sum\limits_{a = 9,10} C_a^{fi} O_a^{fi}\,
,\nn {O_{9(10)}^{fi}} &=\dfrac{\alpha }{4\pi}[\bar s{\gamma ^\mu } P_L b]\,[\bar\ell_f{\gamma _\mu }(\gamma^5)\ell_i] \,,
\label{eq:effHam}
\end{align}
we have
\begin{equation}
C_9^{fi} =  - C_{10}^{fi} = \frac{{ - \sqrt 2 }}{{2{G_F}{V_{tb}}V_{ts}^*}}\frac{\pi }{\alpha }\frac{{\kappa _{2i}^{}\kappa _{3f}^*}}{{{M^2}}}\,.
\end{equation}
The allowed range is given by~\cite{Capdevila:2017bsm}
\begin{equation}
-0.36 (-0.48) \geq  C_{9}^{22}=-C_{10}^{22} \geq (-0.73) -0.87\,,
\end{equation}
at the $2 (1)\,\sigma$ level. In the case of lepton flavor violating $B$ decays, we use the results of Ref.~\cite{Crivellin:2015era} for the analysis of $B\to K^{(*)}\tau\mu$ which currently gives the best experimental limits for $\mu\tau$ final states of~\cite{Lees:2012zz}
\begin{equation}
{\rm{Br}}\left[ {B \to K\tau \mu } \right]_{\rm EXP}\leq4.8\times 10^{-5}\,,
\end{equation}
at $90\%$ confidence level. For our case of $C_9=-C_{10}$ we get
\begin{equation}
{\rm{Br}}\left[ {B \to K\tau \mu } \right] = 1.96 \times {10^{ - 8}}\left( {{{\left| {C_9^{23}} \right|}^2} + {{\left| {C_9^{32}} \right|}^2}} \right)\,.
\end{equation}
Finally, we also get an effect in $B_s\to \tau^+\tau^-$ of
\begin{equation}
{\rm{Br}}\left( {{B_s} \to {\tau ^ + }{\tau ^ - }} \right) = {\rm{Br}}{\left( {{B_s} \to {\tau ^ + }{\tau ^ - }} \right)_{\rm
		SM}}{\left( {1 + \frac{{C_{10}^{33}}}{{C_{10}^{\rm SM}}}} \right)^2}\,,
\end{equation}
with
$	{\rm{Br}}{\left( {B_s \to {\tau ^ + }{\tau ^ - }} \right)_{{\rm{SM}}}} = \left( {7.73 \pm 0.49} \right) \times {10^{ - 7}}$~\cite{Bobeth:2013uxa,Bobeth:2014tza} and $C_{10}^{\rm SM} \approx  - 4.3$~\cite{Bobeth:1999mk,Huber:2005ig}. The current experimental limit is ${\rm{Br}}{\left( {B_s \to {\tau ^ + }{\tau ^ - }} \right)_{{\rm{EXP}}}} \le 6.8 \times {10^{ - 3}}$~\cite{Aaij:2017xqt}.

\subsection{$B_s-\overline{B}_s$ mixing}
With $H=C_1\bar s\gamma^\mu P_L b \bar s\gamma_\mu P_L b$ we get
\begin{equation}
{C_1} = -\frac{{\kappa _{2s}^{}\kappa _{3s}^*\kappa _{2t}^{}\kappa _{3t}^*}}{{16{\pi ^2}}}\left( {\frac{{{D_6}}}{{4M_{LQ}^4}} + {D_2}- \frac{{2{D_4}}}{{M_{LQ}^2}}} \right)\,,
\end{equation}
using unitary gauge. Here $s,t=1-6$ labels the six fermions with the quantum numbers of charged leptons.  Note that after summation over the internal leptons the result is finite due to the GIM-like cancellation originating from our unitary rotation matrices. The standard loop functions $D_x\equiv D_x\left( {{M_{LQ}},{M_{LQ}},{m_s},{m_t}} \right)$ are defined as
\begin{eqnarray}
&\frac{{16{\pi ^2}}}{i}{D_x}\left( {m_1,m_2,m_3,m_4} \right) =  \\
&\int {\frac{{{d^d}k}}{{{{\left( {2\pi } \right)}^d}}}\frac{{{{\left( {{k^2}} \right)}^{x/2}}}}{{\left( {{k^2} - m_1^2} \right)\left( {{k^2} - m_2^2} \right)\left( {{k^2} - m_3^2} \right)\left( {{k^2} - m_4^2} \right)}}}\nonumber
\end{eqnarray}

\begin{figure*}[t]
\begin{center}
\begin{tabular}{cp{7mm}c}
\includegraphics[width=0.39\textwidth]{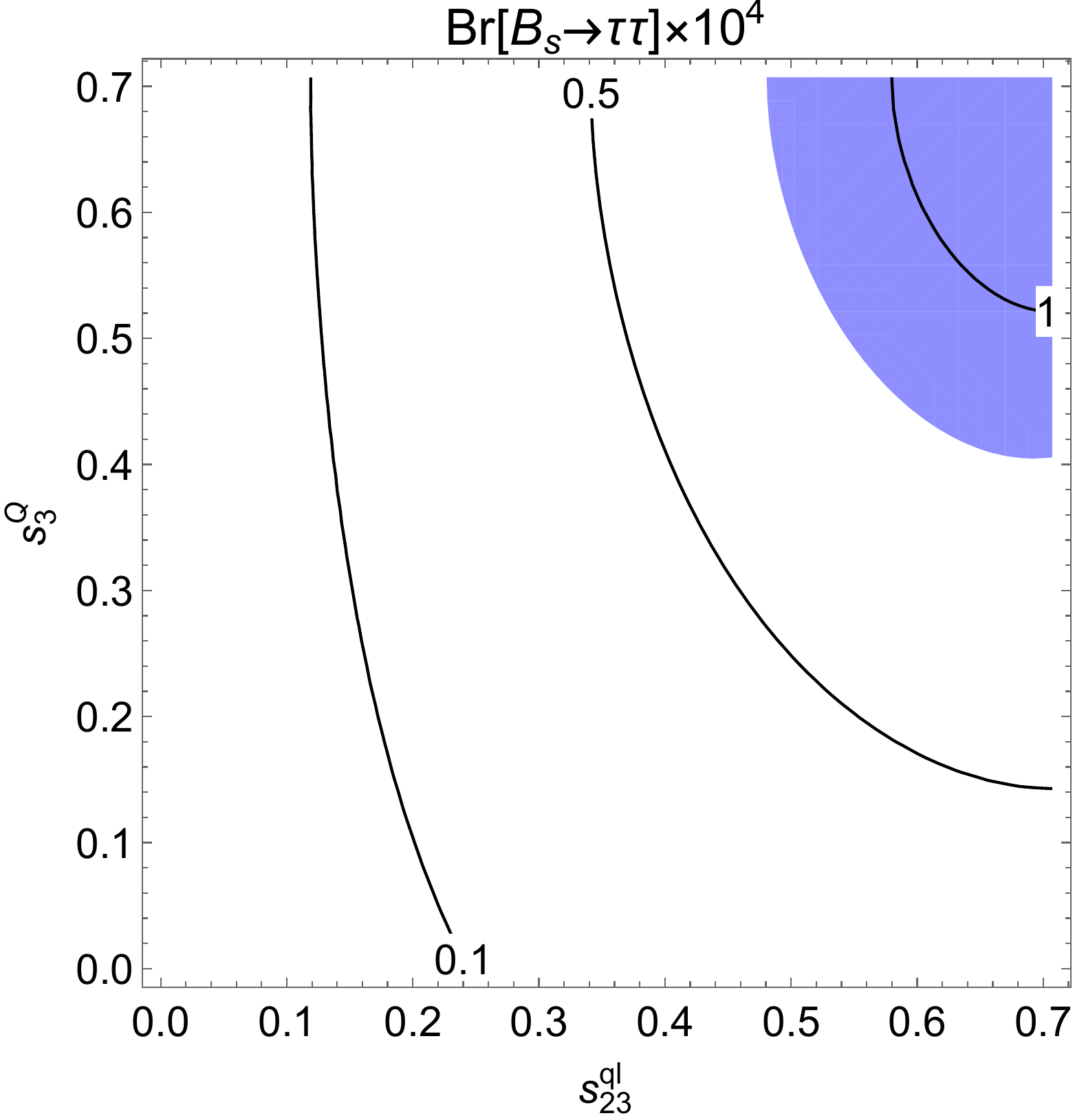}\hspace{1.5cm}
\includegraphics[width=0.53\textwidth]{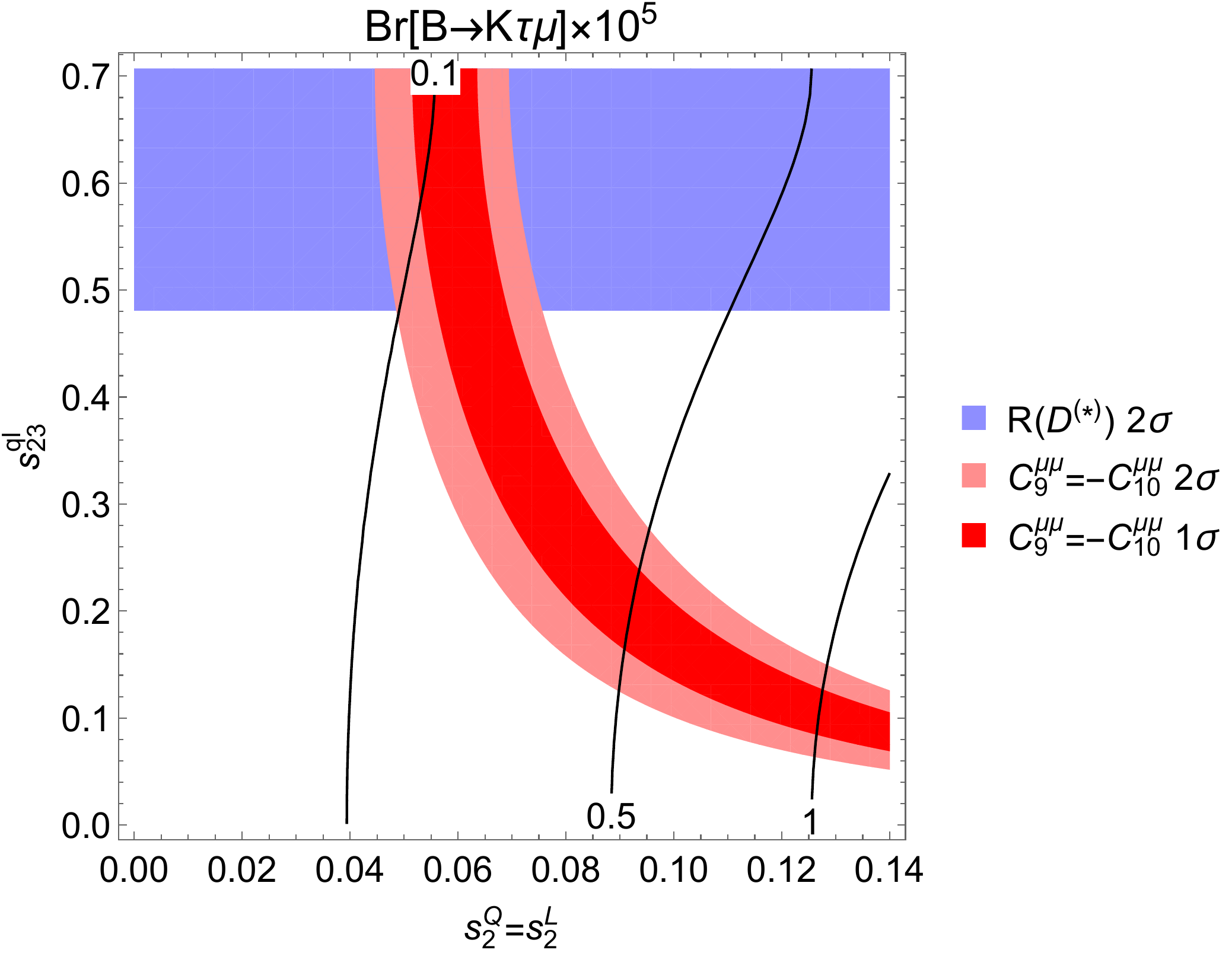}
\end{tabular}
\end{center}
\caption{Left: Allowed regions from $R(D^{(*)})$ for $M_{LQ}=2\,{\rm TeV}$ and $s_3^L=1/\sqrt 2$. Here we used the weighted average for $R(D)$ and $R(D^*)$. The contour lines denote ${\rm Br}(B_s\to\tau^+\tau^-)\times 10^4$. Right: Combined results for $R(D^{(*)})$ and $b\to s\ell^+\ell^-$, and contours for $[{\rm Br}(B\to K\tau^+\mu^-)+{\rm Br}(B\to K\tau^-\mu^+)]/2$. The red region is preferred by the global fit to $b\to s\ell^+\ell^-$ data.}   
\label{Fig1}
\end{figure*}

\section{Phenomenological analysis}
\label{analysis}

Let us first consider $R(D)$ and $R(D^*)$ where the least number of free parameters enters. In order to get a sizable effect, the mixing of $L^L_3$ with the tau lepton $\ell^L_3$ should be large. Assuming it to be maximal (i.e. $M_{33}^{L}=m_{33}^{L}$), we show the regions preferred by $R(D^{(*)})$ in the left plot of Fig.~\ref{Fig1} for $M_{LQ}=2\,$TeV. From this we can see that also the mixing between $Q^L_3$ and $q^L_3$ ($s^Q_3$), as well as the misalignment between the quark and lepton Yukawa couplings of the second and third generations ($s^{q\ell}_{23}$) should be sizable. Our model predicts a significant enhancement of $B_s\to\tau^+\tau^-$~\cite{Alonso:2015sja,Crivellin:2017zlb,Choudhury:2017qyt,Buttazzo:2017ixm} compared to the SM prediction since this process is in our setup mediated at tree-level with order one couplings. 

Let us now turn to the explanation of $b\to s\ell^+\ell^-$ data. Assuming the absence of mixing with leptons of the first generation, we are safe from processes like $\mu\to e\gamma$ or $b\to s\mu e$~\cite{Crivellin:2017dsk} and get the right effect in $R(K)$ and $R(K^*)$. Assuming maximal mixing for the third generation quarks and leptons, we show the preferred region from $b\to s\ell^+\ell^-$ in the right plot of Fig.~\ref{Fig1}. This region overlaps with the one from $R(D^{(*)})$ for small mixing between the second generation fermions ($s_2^{Q,L}$) where the predicted branching ratio for $B\to K\tau\mu$ is automatically compatible with the experimental bounds. However, the predicted rate is still sizable and well within the reach of future measurements. 

So far, we did not specify the absolute mass scale of the vector-like fermions since it did not enter any of the observables. However, for $B_s-\overline{B}_s$ mixing, the masses of the vector-like leptons are crucial. In fact, since we calculated $B_s-\overline{B}_s$ mixing in unitary gauge, the effects of Goldstone bosons are automatically included and therefore the result scales proportional to $(M^L)^2$ (like the SM contribution is proportional to $m_t^2$). Thus, in order to respect the $B_s-\overline{B}_s$ mixing bounds while still accounting for $R(D^{(*)})$, rather light vector-like leptons are required. We checked that the $B_s-\overline{B}_s$ mixing bounds are respected for masses around $500\,$GeV. Since these are third generation leptons, this is compatible with the bounds from direct LHC searches~\cite{Falkowski:2013jya,Dermisek:2014qca}. Anyway, since we only included the effect of the Goldstone bosons and not of physical Higgses in this calculation, this should only be understood as a proof that $B_s-\overline{B}_s$ mixing does not rule out large effects in $R(D^{(*)})$. A more precise prediction would require to specify the Higgs sector explicitly and would be therefore subject to more model dependence.

\section{Including the EW symmetry breaking}
\label{sec:EWSB}
\begin{table}[t]
	\begin{tabular}{c|ccccc}
		{}& ${SU\left( 4 \right)}$&${SU{{\left( 2 \right)}_L}}$&${SU{{\left( 2 \right)}_R}}$
		&${U{{\left( 1 \right)}_{Q}}}$ & ${U{{\left( 1 \right)}_{L}}}$\\
		\hline
		${{X^{L}_i}}$& 4&2&1&0&1\\
		${{Y^{L}_i}}$& 4&2&1&1&0\\
		${{Y^{R}_i}}$& {4}&2&1&0&0\\
		${{X^{R}_i}}$& {4} &1&2&0&1\\
		${{Z^{R}_i}}$& {4}&1&2&1&0\\
		${{Z^{L}_i}}$& 4&1&2&0&0\\
		$\Sigma_{L_1} $& ${\bar 4 \otimes 4}$&1&1&{0} & {-1} \\
		$\Sigma_{L _2}$& ${\bar 4 \otimes 4}$&1&1&{0} & {-1} \\
		$\Sigma_{Q_1}$ & ${\bar 4 \otimes 4}$&1&1&{-1} & {0} \\
		$\Sigma_{Q_2}$ & ${\bar 4 \otimes 4}$&1&1&{-1} & {0} \\
		$\Phi_q $& 1&2&2&0&{-2}\\
		$\Phi_{\ell}$ & 1&2&2&{-2}&0\\
		$\Phi$ &   10&1&3&{-2}&0\\
	\end{tabular}
	\caption{Possible extension of the field content of the model presented in Section \ref{sec:model} accounting for EW breaking.}
	\label{fields-PS}
\end{table}

Finally let us outline a possible UV completion of the Higgs sector which can lead to the desired EW symmetry breaking extending the previously considered particle content to that shown in Table~\ref{fields-PS}. Here we introduced two approximate global symmetries $U{{\left( 1 \right)}_{L,Q}}$ and four Higgs fields $\Sigma_{L_a}$ and $\Sigma_{Q_a}$, $a=1,2$, which generate the vector-like masses for $(Q_L,~Q'_R)$, $(L_L,~L'_R)$, $(Q_R,~Q'_L)$, and $(L_R,~L'_L)$ as in Section \ref{sec:model}. $\Phi_q$ and $\Phi_{\ell}$ will generate the SM fermion masses and mixing. $\Phi$ will break the $SU(4)\times SU(2)_L\times SU(2)_R$ symmetry down to the SM gauge symmetry and gives Majorana masses to the right-handed neutrinos.
We choose the vevs for $\Sigma_{L_a}$, $\Sigma_{Q_a}$, $a=1,2$,
$\Phi_q$, and $\Phi_{\ell}$ as follows
\begin{eqnarray}
\Sigma_{L_a} &=& {\rm diag}\left(0,~0,~0,~v_{L_a}\right),~\nonumber \\
\Sigma_{Q_a} &=& {\rm diag}\left(v_{Q_a},~v_{Q_a},~v_{Q_a},~0\right),~\nonumber \\
\Phi_q &=& {\rm diag}\left(v_d,~v_u\right),~\nonumber \\
\Phi_{\ell} &=& {\rm diag}\left(v_{\ell},~v_{\nu}\right).~
\end{eqnarray}
The Lagrangian for vector-like particle masses and SM fermion masses are
\begin{eqnarray}
-{\cal L} &=& y_{ij}^{L_a} \bar X_i^L Y_j^R \Sigma_{L_a} + y_{ij}^{Q_a} \bar Y_i^L Y_j^R \Sigma_{Q_a}
\nonumber \\ &&
+ y_{ij}^{L'_a} \bar X_i^R Z_j^{L}  \Sigma_{L_a}
+ y_{ij}^{Q_a'} \bar Z_i^R Z_j^L \Sigma_{Q_a} \nonumber \\ &&
+ y_{ij}^q \bar X_i^L X_j^R\Phi_q + y_{ij}^{\ell} \bar Y_i^L Z_j^R \Phi_{\ell} 
\nonumber \\ &&
+ y_{ij}^N \bar Z_i^R Z_j^R \Phi + h.c.
\label{Yukawa-Coupling-P}
\end{eqnarray}
Thus, without further contributions, the up-type quark mass matrix would be proportional to 
the down-type quark mass matrix. To generate the couplings between the SM left-handed fermions and massive gauge bosons and solve the quark mass problem, we introduce the following terms which violate the $U(1)_Q\times U(1)_L$ global symmetries
\begin{eqnarray}
-{\cal L} &=& h_{ij}^{Q} \bar X_i^L Y_j^R \Sigma_Q + h_{ij}^{L} \bar Y_i^L Y_j^R \Sigma_L
\nonumber \\ &&
+ h_{ij}^q \bar X_i^L X_j^R\Phi_{\ell}
+ h_{ij}^{\ell} \bar Y_i^L Z_j^R \Phi_{q} +h.c.\,,
\label{Yukawa-Coupling-V}
\end{eqnarray}
where we expect the above Yukawa couplings to be relatively small
compared to these in Eq.~(\ref{Yukawa-Coupling-P}) due to 
$U(1)_Q\times U(1)_L$ global symmetry breaking.
Notice that the first line gives rise to mass terms of the form given in Eq.~(\ref{Lmass}).
We defer to future publication a complete analysis of this and other possible completions of our setup, including the flavor effects induced by the extra Higgses. 

\section{Conclusions and outlook}\label{conclusions}
In this article we presented a renormalizable phenomenologically valid TeV scale model of a vector leptoquarks with flavor dependent couplings. The model is an extension of the PS model obtained by adding three generations of vector-like fermions which are in fundamental representations of $SU(4)$. Our model can successfully address the observed deviations from the SM predictions in semi-leptonic $B$ decays ($R(D^{(*)})$ as well as in $b\to s\ell^+\ell^-$ transitions) and easily account for the anomaly in the anomalous magnetic moment of the muon too, as we comment below. An explanation of $R(D^{(*)})$ predicts a significant enhancement of $B_s\to\tau^+\tau^-$ and once also $b\to s\ell^+\ell^-$ is included, sizable rates for $b\to\tau\mu$ processes must occur. Also bounds from $B_s-\overline{B}_s$ mixing are respected for not too heavy vector-like leptons.

The longstanding anomaly in the anomalous magnetic moment of the muon (at the $3\,\sigma$ level) might also be related to the $B$-physics anomalies~\cite{Crivellin:2015hha,Belanger:2015nma,Bauer:2015knc,Altmannshofer:2016oaq,Das:2016vkr,Chen:2016dip,Arnan:2016cpy,ColuccioLeskow:2016dox,Crivellin:2017zlb,DiChiara:2017cjq}. Our model can in principle explain this discrepancy, as long as one introduces couplings of heavy down-type quarks to the SM Higgs doublet. Following \eq{RH-embedding}, we call the vector like fermions with the same quantum numbers as right-handed down-quarks $D^R_{i}$ and ${D'}^{L}_i$. Then the coupling to the Higgs is given by $\lambda_{ii}\bar D'^L_i \phi D^R_{i}$ where our coupling of interest is $\lambda_{22}$. For vector-like quarks and leptoquark of equal mass $M$, the resulting contribution is $\delta {a_\mu } \approx \frac{{{m_\mu }}}{{4{\pi ^2}}}\frac{{2v\lambda_{22} }}{{{M^2}}}$. Since $\lambda_{22}$ does not enter in other observables discussed to far, it can simply be adjusted to account for the anomaly. Note that this is possible for natural values (smaller than 0.1). 

\medskip
\noindent {\it Note added} --- During the completion of the article a model of vector leptoquarks also based on the PS group was presented in Ref.~\cite{DiLuzio:2017vat} (for an earlier attempt see also Ref.~\cite{Assad:2017iib}) . While in our model all fermions transform in the fundamental of $SU(4)$, in Ref.~\cite{DiLuzio:2017vat} the gauge group contains another $SU(3)$ factor and the SM-like fermions are singlets of $SU(4)$. 
Also note that in our model only the leptoquark has flavor non-universal couplings while in the model of Ref.~\cite{DiLuzio:2017vat} also the heavy $Z'$ and the heavy gluons in general acquire flavor violating couplings.

\medskip
%
\noindent {\it Acknowledgments} --- {The work of A.C.~is supported by an Ambizione Grant of the Swiss National Science Foundation (PZ00P2\_154834). This research was supported in part by the Projects 11475238 and 11647601 supported by the National Natural Science Foundation of China, and by the Key Research Program of Frontier Science, CAS. A.C.~thanks Dario M\"uller for useful comments on the manuscript.
We are especially grateful to Luca Di Luzio, Admir Greljo, and Marco Nardecchia for discussions and for pointing out that in an earlier version of the paper lepton and quark rotations were not independent, such that the desired pattern for the masses could not be achieved. }

\bibliography{BIB}

\begin{thebibliography}{70}
\expandafter\ifx\csname natexlab\endcsname\relax\def\natexlab#1{#1}\fi
\expandafter\ifx\csname bibnamefont\endcsname\relax
  \def\bibnamefont#1{#1}\fi
\expandafter\ifx\csname bibfnamefont\endcsname\relax
  \def\bibfnamefont#1{#1}\fi
\expandafter\ifx\csname citenamefont\endcsname\relax
  \def\citenamefont#1{#1}\fi
\expandafter\ifx\csname url\endcsname\relax
  \def\url#1{\texttt{#1}}\fi
\expandafter\ifx\csname urlprefix\endcsname\relax\def\urlprefix{URL }\fi
\providecommand{\bibinfo}[2]{#2}
\providecommand{\eprint}[2][]{\url{#2}}

\bibitem[{\citenamefont{Capdevila et~al.}(2017)\citenamefont{Capdevila,
  Crivellin, Descotes-Genon, Matias, and Virto}}]{Capdevila:2017bsm}
\bibinfo{author}{\bibfnamefont{B.}~\bibnamefont{Capdevila}},
  \bibinfo{author}{\bibfnamefont{A.}~\bibnamefont{Crivellin}},
  \bibinfo{author}{\bibfnamefont{S.}~\bibnamefont{Descotes-Genon}},
  \bibinfo{author}{\bibfnamefont{J.}~\bibnamefont{Matias}}, \bibnamefont{and}
  \bibinfo{author}{\bibfnamefont{J.}~\bibnamefont{Virto}}
  (\bibinfo{year}{2017}), \eprint{1704.05340}.

\bibitem[{\citenamefont{Altmannshofer et~al.}(2017)\citenamefont{Altmannshofer,
  Stangl, and Straub}}]{Altmannshofer:2017yso}
\bibinfo{author}{\bibfnamefont{W.}~\bibnamefont{Altmannshofer}},
  \bibinfo{author}{\bibfnamefont{P.}~\bibnamefont{Stangl}}, \bibnamefont{and}
  \bibinfo{author}{\bibfnamefont{D.~M.} \bibnamefont{Straub}}
  (\bibinfo{year}{2017}), \eprint{1704.05435}.

\bibitem[{\citenamefont{D'Amico et~al.}(2017)\citenamefont{D'Amico, Nardecchia,
  Panci, Sannino, Strumia, Torre, and Urbano}}]{DAmico:2017mtc}
\bibinfo{author}{\bibfnamefont{G.}~\bibnamefont{D'Amico}},
  \bibinfo{author}{\bibfnamefont{M.}~\bibnamefont{Nardecchia}},
  \bibinfo{author}{\bibfnamefont{P.}~\bibnamefont{Panci}},
  \bibinfo{author}{\bibfnamefont{F.}~\bibnamefont{Sannino}},
  \bibinfo{author}{\bibfnamefont{A.}~\bibnamefont{Strumia}},
  \bibinfo{author}{\bibfnamefont{R.}~\bibnamefont{Torre}}, \bibnamefont{and}
  \bibinfo{author}{\bibfnamefont{A.}~\bibnamefont{Urbano}}
  (\bibinfo{year}{2017}), \eprint{1704.05438}.

\bibitem[{\citenamefont{Geng et~al.}(2017)\citenamefont{Geng, Grinstein, Jager,
  Martin~Camalich, Ren, and Shi}}]{Geng:2017svp}
\bibinfo{author}{\bibfnamefont{L.-S.} \bibnamefont{Geng}},
  \bibinfo{author}{\bibfnamefont{B.}~\bibnamefont{Grinstein}},
  \bibinfo{author}{\bibfnamefont{S.}~\bibnamefont{Jager}},
  \bibinfo{author}{\bibfnamefont{J.}~\bibnamefont{Martin~Camalich}},
  \bibinfo{author}{\bibfnamefont{X.-L.} \bibnamefont{Ren}}, \bibnamefont{and}
  \bibinfo{author}{\bibfnamefont{R.-X.} \bibnamefont{Shi}}
  (\bibinfo{year}{2017}), \eprint{1704.05446}.

\bibitem[{\citenamefont{Ciuchini et~al.}(2017)\citenamefont{Ciuchini, Coutinho,
  Fedele, Franco, Paul, Silvestrini, and Valli}}]{Ciuchini:2017mik}
\bibinfo{author}{\bibfnamefont{M.}~\bibnamefont{Ciuchini}},
  \bibinfo{author}{\bibfnamefont{A.~M.} \bibnamefont{Coutinho}},
  \bibinfo{author}{\bibfnamefont{M.}~\bibnamefont{Fedele}},
  \bibinfo{author}{\bibfnamefont{E.}~\bibnamefont{Franco}},
  \bibinfo{author}{\bibfnamefont{A.}~\bibnamefont{Paul}},
  \bibinfo{author}{\bibfnamefont{L.}~\bibnamefont{Silvestrini}},
  \bibnamefont{and} \bibinfo{author}{\bibfnamefont{M.}~\bibnamefont{Valli}}
  (\bibinfo{year}{2017}), \eprint{1704.05447}.

\bibitem[{\citenamefont{Hiller and Nisandzic}(2017)}]{Hiller:2017bzc}
\bibinfo{author}{\bibfnamefont{G.}~\bibnamefont{Hiller}} \bibnamefont{and}
  \bibinfo{author}{\bibfnamefont{I.}~\bibnamefont{Nisandzic}}
  (\bibinfo{year}{2017}), \eprint{1704.05444}.

\bibitem[{\citenamefont{Alok et~al.}(2017{\natexlab{a}})\citenamefont{Alok,
  Bhattacharya, Datta, Kumar, Kumar, and London}}]{Alok:2017sui}
\bibinfo{author}{\bibfnamefont{A.~K.} \bibnamefont{Alok}},
  \bibinfo{author}{\bibfnamefont{B.}~\bibnamefont{Bhattacharya}},
  \bibinfo{author}{\bibfnamefont{A.}~\bibnamefont{Datta}},
  \bibinfo{author}{\bibfnamefont{D.}~\bibnamefont{Kumar}},
  \bibinfo{author}{\bibfnamefont{J.}~\bibnamefont{Kumar}}, \bibnamefont{and}
  \bibinfo{author}{\bibfnamefont{D.}~\bibnamefont{London}}
  (\bibinfo{year}{2017}{\natexlab{a}}), \eprint{1704.07397}.

\bibitem[{\citenamefont{Hurth et~al.}(2017)\citenamefont{Hurth, Mahmoudi,
  Martinez~Santos, and Neshatpour}}]{Hurth:2017hxg}
\bibinfo{author}{\bibfnamefont{T.}~\bibnamefont{Hurth}},
  \bibinfo{author}{\bibfnamefont{F.}~\bibnamefont{Mahmoudi}},
  \bibinfo{author}{\bibfnamefont{D.}~\bibnamefont{Martinez~Santos}},
  \bibnamefont{and}
  \bibinfo{author}{\bibfnamefont{S.}~\bibnamefont{Neshatpour}}
  (\bibinfo{year}{2017}), \eprint{1705.06274}.

\bibitem[{\citenamefont{Amhis et~al.}(2016)}]{Amhis:2016xyh}
\bibinfo{author}{\bibfnamefont{Y.}~\bibnamefont{Amhis}} \bibnamefont{et~al.}
  (\bibinfo{year}{2016}), \eprint{1612.07233}.

\bibitem[{\citenamefont{Aaij et~al.}(2014)}]{Aaij:2014ora}
\bibinfo{author}{\bibfnamefont{R.}~\bibnamefont{Aaij}} \bibnamefont{et~al.}
  (\bibinfo{collaboration}{LHCb collaboration}),
  \bibinfo{journal}{Phys.Rev.Lett.} \textbf{\bibinfo{volume}{113}},
  \bibinfo{pages}{151601} (\bibinfo{year}{2014}), \eprint{1406.6482}.

\bibitem[{\citenamefont{Aaij et~al.}(2017{\natexlab{a}})}]{Aaij:2017vbb}
\bibinfo{author}{\bibfnamefont{R.}~\bibnamefont{Aaij}} \bibnamefont{et~al.}
  (\bibinfo{collaboration}{LHCb}) (\bibinfo{year}{2017}{\natexlab{a}}),
  \eprint{1705.05802}.

\bibitem[{\citenamefont{Bhattacharya et~al.}(2015)\citenamefont{Bhattacharya,
  Datta, London, and Shivashankara}}]{Bhattacharya:2014wla}
\bibinfo{author}{\bibfnamefont{B.}~\bibnamefont{Bhattacharya}},
  \bibinfo{author}{\bibfnamefont{A.}~\bibnamefont{Datta}},
  \bibinfo{author}{\bibfnamefont{D.}~\bibnamefont{London}}, \bibnamefont{and}
  \bibinfo{author}{\bibfnamefont{S.}~\bibnamefont{Shivashankara}},
  \bibinfo{journal}{Phys. Lett.} \textbf{\bibinfo{volume}{B742}},
  \bibinfo{pages}{370} (\bibinfo{year}{2015}), \eprint{1412.7164}.

\bibitem[{\citenamefont{Calibbi et~al.}(2015)\citenamefont{Calibbi, Crivellin,
  and Ota}}]{Calibbi:2015kma}
\bibinfo{author}{\bibfnamefont{L.}~\bibnamefont{Calibbi}},
  \bibinfo{author}{\bibfnamefont{A.}~\bibnamefont{Crivellin}},
  \bibnamefont{and} \bibinfo{author}{\bibfnamefont{T.}~\bibnamefont{Ota}},
  \bibinfo{journal}{Phys. Rev. Lett.} \textbf{\bibinfo{volume}{115}},
  \bibinfo{pages}{181801} (\bibinfo{year}{2015}), \eprint{1506.02661}.

\bibitem[{\citenamefont{Fajfer and Košnik}(2016)}]{Fajfer:2015ycq}
\bibinfo{author}{\bibfnamefont{S.}~\bibnamefont{Fajfer}} \bibnamefont{and}
  \bibinfo{author}{\bibfnamefont{N.}~\bibnamefont{Košnik}},
  \bibinfo{journal}{Phys. Lett.} \textbf{\bibinfo{volume}{B755}},
  \bibinfo{pages}{270} (\bibinfo{year}{2016}), \eprint{1511.06024}.

\bibitem[{\citenamefont{Greljo et~al.}(2015)\citenamefont{Greljo, Isidori, and
  Marzocca}}]{Greljo:2015mma}
\bibinfo{author}{\bibfnamefont{A.}~\bibnamefont{Greljo}},
  \bibinfo{author}{\bibfnamefont{G.}~\bibnamefont{Isidori}}, \bibnamefont{and}
  \bibinfo{author}{\bibfnamefont{D.}~\bibnamefont{Marzocca}},
  \bibinfo{journal}{JHEP} \textbf{\bibinfo{volume}{07}}, \bibinfo{pages}{142}
  (\bibinfo{year}{2015}), \eprint{1506.01705}.

\bibitem[{\citenamefont{Barbieri et~al.}(2016)\citenamefont{Barbieri, Isidori,
  Pattori, and Senia}}]{Barbieri:2015yvd}
\bibinfo{author}{\bibfnamefont{R.}~\bibnamefont{Barbieri}},
  \bibinfo{author}{\bibfnamefont{G.}~\bibnamefont{Isidori}},
  \bibinfo{author}{\bibfnamefont{A.}~\bibnamefont{Pattori}}, \bibnamefont{and}
  \bibinfo{author}{\bibfnamefont{F.}~\bibnamefont{Senia}},
  \bibinfo{journal}{Eur. Phys. J.} \textbf{\bibinfo{volume}{C76}},
  \bibinfo{pages}{67} (\bibinfo{year}{2016}), \eprint{1512.01560}.

\bibitem[{\citenamefont{Bauer and Neubert}(2016)}]{Bauer:2015knc}
\bibinfo{author}{\bibfnamefont{M.}~\bibnamefont{Bauer}} \bibnamefont{and}
  \bibinfo{author}{\bibfnamefont{M.}~\bibnamefont{Neubert}},
  \bibinfo{journal}{Phys. Rev. Lett.} \textbf{\bibinfo{volume}{116}},
  \bibinfo{pages}{141802} (\bibinfo{year}{2016}), \eprint{1511.01900}.

\bibitem[{\citenamefont{Boucenna et~al.}(2016)\citenamefont{Boucenna, Celis,
  Fuentes-Martin, Vicente, and Virto}}]{Boucenna:2016qad}
\bibinfo{author}{\bibfnamefont{S.~M.} \bibnamefont{Boucenna}},
  \bibinfo{author}{\bibfnamefont{A.}~\bibnamefont{Celis}},
  \bibinfo{author}{\bibfnamefont{J.}~\bibnamefont{Fuentes-Martin}},
  \bibinfo{author}{\bibfnamefont{A.}~\bibnamefont{Vicente}}, \bibnamefont{and}
  \bibinfo{author}{\bibfnamefont{J.}~\bibnamefont{Virto}},
  \bibinfo{journal}{JHEP} \textbf{\bibinfo{volume}{12}}, \bibinfo{pages}{059}
  (\bibinfo{year}{2016}), \eprint{1608.01349}.

\bibitem[{\citenamefont{Das et~al.}(2016)\citenamefont{Das, Hati, Kumar, and
  Mahajan}}]{Das:2016vkr}
\bibinfo{author}{\bibfnamefont{D.}~\bibnamefont{Das}},
  \bibinfo{author}{\bibfnamefont{C.}~\bibnamefont{Hati}},
  \bibinfo{author}{\bibfnamefont{G.}~\bibnamefont{Kumar}}, \bibnamefont{and}
  \bibinfo{author}{\bibfnamefont{N.}~\bibnamefont{Mahajan}},
  \bibinfo{journal}{Phys. Rev.} \textbf{\bibinfo{volume}{D94}},
  \bibinfo{pages}{055034} (\bibinfo{year}{2016}), \eprint{1605.06313}.

\bibitem[{\citenamefont{Bečirević et~al.}(2016)\citenamefont{Bečirević,
  Fajfer, Košnik, and Sumensari}}]{Becirevic:2016yqi}
\bibinfo{author}{\bibfnamefont{D.}~\bibnamefont{Bečirević}},
  \bibinfo{author}{\bibfnamefont{S.}~\bibnamefont{Fajfer}},
  \bibinfo{author}{\bibfnamefont{N.}~\bibnamefont{Košnik}}, \bibnamefont{and}
  \bibinfo{author}{\bibfnamefont{O.}~\bibnamefont{Sumensari}},
  \bibinfo{journal}{Phys. Rev.} \textbf{\bibinfo{volume}{D94}},
  \bibinfo{pages}{115021} (\bibinfo{year}{2016}), \eprint{1608.08501}.

\bibitem[{\citenamefont{Sahoo et~al.}(2017)\citenamefont{Sahoo, Mohanta, and
  Giri}}]{Sahoo:2016pet}
\bibinfo{author}{\bibfnamefont{S.}~\bibnamefont{Sahoo}},
  \bibinfo{author}{\bibfnamefont{R.}~\bibnamefont{Mohanta}}, \bibnamefont{and}
  \bibinfo{author}{\bibfnamefont{A.~K.} \bibnamefont{Giri}},
  \bibinfo{journal}{Phys. Rev.} \textbf{\bibinfo{volume}{D95}},
  \bibinfo{pages}{035027} (\bibinfo{year}{2017}), \eprint{1609.04367}.

\bibitem[{\citenamefont{Bhattacharya et~al.}(2017)\citenamefont{Bhattacharya,
  Datta, Guevin, London, and Watanabe}}]{Bhattacharya:2016mcc}
\bibinfo{author}{\bibfnamefont{B.}~\bibnamefont{Bhattacharya}},
  \bibinfo{author}{\bibfnamefont{A.}~\bibnamefont{Datta}},
  \bibinfo{author}{\bibfnamefont{J.-P.} \bibnamefont{Guevin}},
  \bibinfo{author}{\bibfnamefont{D.}~\bibnamefont{London}}, \bibnamefont{and}
  \bibinfo{author}{\bibfnamefont{R.}~\bibnamefont{Watanabe}},
  \bibinfo{journal}{JHEP} \textbf{\bibinfo{volume}{01}}, \bibinfo{pages}{015}
  (\bibinfo{year}{2017}), \eprint{1609.09078}.

\bibitem[{\citenamefont{Barbieri et~al.}(2017)\citenamefont{Barbieri, Murphy,
  and Senia}}]{Barbieri:2016las}
\bibinfo{author}{\bibfnamefont{R.}~\bibnamefont{Barbieri}},
  \bibinfo{author}{\bibfnamefont{C.~W.} \bibnamefont{Murphy}},
  \bibnamefont{and} \bibinfo{author}{\bibfnamefont{F.}~\bibnamefont{Senia}},
  \bibinfo{journal}{Eur. Phys. J.} \textbf{\bibinfo{volume}{C77}},
  \bibinfo{pages}{8} (\bibinfo{year}{2017}), \eprint{1611.04930}.

\bibitem[{\citenamefont{Alok et~al.}(2017{\natexlab{b}})\citenamefont{Alok,
  Kumar, Kumar, and Sharma}}]{Alok:2017jaf}
\bibinfo{author}{\bibfnamefont{A.~K.} \bibnamefont{Alok}},
  \bibinfo{author}{\bibfnamefont{D.}~\bibnamefont{Kumar}},
  \bibinfo{author}{\bibfnamefont{J.}~\bibnamefont{Kumar}}, \bibnamefont{and}
  \bibinfo{author}{\bibfnamefont{R.}~\bibnamefont{Sharma}}
  (\bibinfo{year}{2017}{\natexlab{b}}), \eprint{1704.07347}.

\bibitem[{\citenamefont{Crivellin
  et~al.}(2017{\natexlab{a}})\citenamefont{Crivellin, Mueller, and
  Ota}}]{Crivellin:2017zlb}
\bibinfo{author}{\bibfnamefont{A.}~\bibnamefont{Crivellin}},
  \bibinfo{author}{\bibfnamefont{D.}~\bibnamefont{Mueller}}, \bibnamefont{and}
  \bibinfo{author}{\bibfnamefont{T.}~\bibnamefont{Ota}}
  (\bibinfo{year}{2017}{\natexlab{a}}), \eprint{1703.09226}.

\bibitem[{\citenamefont{Chen et~al.}(2017)\citenamefont{Chen, Nomura, and
  Okada}}]{Chen:2017hir}
\bibinfo{author}{\bibfnamefont{C.-H.} \bibnamefont{Chen}},
  \bibinfo{author}{\bibfnamefont{T.}~\bibnamefont{Nomura}}, \bibnamefont{and}
  \bibinfo{author}{\bibfnamefont{H.}~\bibnamefont{Okada}}
  (\bibinfo{year}{2017}), \eprint{1703.03251}.

\bibitem[{\citenamefont{Doršner et~al.}(2017)\citenamefont{Doršner, Fajfer,
  Faroughy, and Košnik}}]{Dorsner:2017ufx}
\bibinfo{author}{\bibfnamefont{I.}~\bibnamefont{Doršner}},
  \bibinfo{author}{\bibfnamefont{S.}~\bibnamefont{Fajfer}},
  \bibinfo{author}{\bibfnamefont{D.~A.} \bibnamefont{Faroughy}},
  \bibnamefont{and} \bibinfo{author}{\bibfnamefont{N.}~\bibnamefont{Košnik}}
  (\bibinfo{year}{2017}), \eprint{1706.07779}.

\bibitem[{\citenamefont{Buttazzo et~al.}(2017)\citenamefont{Buttazzo, Greljo,
  Isidori, and Marzocca}}]{Buttazzo:2017ixm}
\bibinfo{author}{\bibfnamefont{D.}~\bibnamefont{Buttazzo}},
  \bibinfo{author}{\bibfnamefont{A.}~\bibnamefont{Greljo}},
  \bibinfo{author}{\bibfnamefont{G.}~\bibnamefont{Isidori}}, \bibnamefont{and}
  \bibinfo{author}{\bibfnamefont{D.}~\bibnamefont{Marzocca}}
  (\bibinfo{year}{2017}), \eprint{1706.07808}.

\bibitem[{\citenamefont{Faroughy et~al.}(2017)\citenamefont{Faroughy, Greljo,
  and Kamenik}}]{Faroughy:2016osc}
\bibinfo{author}{\bibfnamefont{D.~A.} \bibnamefont{Faroughy}},
  \bibinfo{author}{\bibfnamefont{A.}~\bibnamefont{Greljo}}, \bibnamefont{and}
  \bibinfo{author}{\bibfnamefont{J.}~\bibnamefont{Kamenik}},
  \bibinfo{journal}{Phys. Lett.} \textbf{\bibinfo{volume}{B764}},
  \bibinfo{pages}{126} (\bibinfo{year}{2017}), \eprint{1609.07138}.

\bibitem[{\citenamefont{Feruglio et~al.}(2017)\citenamefont{Feruglio, Paradisi,
  and Pattori}}]{Feruglio:2016gvd}
\bibinfo{author}{\bibfnamefont{F.}~\bibnamefont{Feruglio}},
  \bibinfo{author}{\bibfnamefont{P.}~\bibnamefont{Paradisi}}, \bibnamefont{and}
  \bibinfo{author}{\bibfnamefont{A.}~\bibnamefont{Pattori}},
  \bibinfo{journal}{Phys. Rev. Lett.} \textbf{\bibinfo{volume}{118}},
  \bibinfo{pages}{011801} (\bibinfo{year}{2017}), \eprint{1606.00524}.

\bibitem[{\citenamefont{Crivellin et~al.}(2012)\citenamefont{Crivellin, Greub,
  and Kokulu}}]{Crivellin:2012ye}
\bibinfo{author}{\bibfnamefont{A.}~\bibnamefont{Crivellin}},
  \bibinfo{author}{\bibfnamefont{C.}~\bibnamefont{Greub}}, \bibnamefont{and}
  \bibinfo{author}{\bibfnamefont{A.}~\bibnamefont{Kokulu}},
  \bibinfo{journal}{Phys.Rev.} \textbf{\bibinfo{volume}{D86}},
  \bibinfo{pages}{054014} (\bibinfo{year}{2012}), \eprint{1206.2634}.

\bibitem[{\citenamefont{Tanaka and Watanabe}(2013)}]{Tanaka:2012nw}
\bibinfo{author}{\bibfnamefont{M.}~\bibnamefont{Tanaka}} \bibnamefont{and}
  \bibinfo{author}{\bibfnamefont{R.}~\bibnamefont{Watanabe}},
  \bibinfo{journal}{Phys. Rev.} \textbf{\bibinfo{volume}{D87}},
  \bibinfo{pages}{034028} (\bibinfo{year}{2013}), \eprint{1212.1878}.

\bibitem[{\citenamefont{Celis et~al.}(2013)\citenamefont{Celis, Jung, Li, and
  Pich}}]{Celis:2012dk}
\bibinfo{author}{\bibfnamefont{A.}~\bibnamefont{Celis}},
  \bibinfo{author}{\bibfnamefont{M.}~\bibnamefont{Jung}},
  \bibinfo{author}{\bibfnamefont{X.-Q.} \bibnamefont{Li}}, \bibnamefont{and}
  \bibinfo{author}{\bibfnamefont{A.}~\bibnamefont{Pich}},
  \bibinfo{journal}{JHEP} \textbf{\bibinfo{volume}{1301}}, \bibinfo{pages}{054}
  (\bibinfo{year}{2013}), \eprint{1210.8443}.

\bibitem[{\citenamefont{Crivellin et~al.}(2013)\citenamefont{Crivellin, Kokulu,
  and Greub}}]{Crivellin:2013wna}
\bibinfo{author}{\bibfnamefont{A.}~\bibnamefont{Crivellin}},
  \bibinfo{author}{\bibfnamefont{A.}~\bibnamefont{Kokulu}}, \bibnamefont{and}
  \bibinfo{author}{\bibfnamefont{C.}~\bibnamefont{Greub}},
  \bibinfo{journal}{Phys.Rev.} \textbf{\bibinfo{volume}{D87}},
  \bibinfo{pages}{094031} (\bibinfo{year}{2013}), \eprint{1303.5877}.

\bibitem[{\citenamefont{Crivellin
  et~al.}(2016{\natexlab{a}})\citenamefont{Crivellin, Heeck, and
  Stoffer}}]{Crivellin:2015hha}
\bibinfo{author}{\bibfnamefont{A.}~\bibnamefont{Crivellin}},
  \bibinfo{author}{\bibfnamefont{J.}~\bibnamefont{Heeck}}, \bibnamefont{and}
  \bibinfo{author}{\bibfnamefont{P.}~\bibnamefont{Stoffer}},
  \bibinfo{journal}{Phys. Rev. Lett.} \textbf{\bibinfo{volume}{116}},
  \bibinfo{pages}{081801} (\bibinfo{year}{2016}{\natexlab{a}}),
  \eprint{1507.07567}.

\bibitem[{\citenamefont{Chen and Nomura}(2017)}]{Chen:2017eby}
\bibinfo{author}{\bibfnamefont{C.-H.} \bibnamefont{Chen}} \bibnamefont{and}
  \bibinfo{author}{\bibfnamefont{T.}~\bibnamefont{Nomura}}
  (\bibinfo{year}{2017}), \eprint{1703.03646}.

\bibitem[{\citenamefont{Freytsis et~al.}(2015)\citenamefont{Freytsis, Ligeti,
  and Ruderman}}]{Freytsis:2015qca}
\bibinfo{author}{\bibfnamefont{M.}~\bibnamefont{Freytsis}},
  \bibinfo{author}{\bibfnamefont{Z.}~\bibnamefont{Ligeti}}, \bibnamefont{and}
  \bibinfo{author}{\bibfnamefont{J.~T.} \bibnamefont{Ruderman}},
  \bibinfo{journal}{Phys. Rev.} \textbf{\bibinfo{volume}{D92}},
  \bibinfo{pages}{054018} (\bibinfo{year}{2015}), \eprint{1506.08896}.

\bibitem[{\citenamefont{Celis et~al.}(2017)\citenamefont{Celis, Jung, Li, and
  Pich}}]{Celis:2016azn}
\bibinfo{author}{\bibfnamefont{A.}~\bibnamefont{Celis}},
  \bibinfo{author}{\bibfnamefont{M.}~\bibnamefont{Jung}},
  \bibinfo{author}{\bibfnamefont{X.-Q.} \bibnamefont{Li}}, \bibnamefont{and}
  \bibinfo{author}{\bibfnamefont{A.}~\bibnamefont{Pich}},
  \bibinfo{journal}{Phys. Lett.} \textbf{\bibinfo{volume}{B771}},
  \bibinfo{pages}{168} (\bibinfo{year}{2017}), \eprint{1612.07757}.

\bibitem[{\citenamefont{Ivanov et~al.}(2017)\citenamefont{Ivanov, Koerner, and
  Tran}}]{Ivanov:2017mrj}
\bibinfo{author}{\bibfnamefont{M.~A.} \bibnamefont{Ivanov}},
  \bibinfo{author}{\bibfnamefont{J.~G.} \bibnamefont{Koerner}},
  \bibnamefont{and} \bibinfo{author}{\bibfnamefont{C.-T.} \bibnamefont{Tran}},
  \bibinfo{journal}{Phys. Rev.} \textbf{\bibinfo{volume}{D95}},
  \bibinfo{pages}{036021} (\bibinfo{year}{2017}), \eprint{1701.02937}.

\bibitem[{\citenamefont{Li et~al.}(2016)\citenamefont{Li, Yang, and
  Zhang}}]{Li:2016vvp}
\bibinfo{author}{\bibfnamefont{X.-Q.} \bibnamefont{Li}},
  \bibinfo{author}{\bibfnamefont{Y.-D.} \bibnamefont{Yang}}, \bibnamefont{and}
  \bibinfo{author}{\bibfnamefont{X.}~\bibnamefont{Zhang}},
  \bibinfo{journal}{JHEP} \textbf{\bibinfo{volume}{08}}, \bibinfo{pages}{054}
  (\bibinfo{year}{2016}), \eprint{1605.09308}.

\bibitem[{\citenamefont{Alonso et~al.}(2017)\citenamefont{Alonso, Grinstein,
  and Martin~Camalich}}]{Alonso:2016oyd}
\bibinfo{author}{\bibfnamefont{R.}~\bibnamefont{Alonso}},
  \bibinfo{author}{\bibfnamefont{B.}~\bibnamefont{Grinstein}},
  \bibnamefont{and}
  \bibinfo{author}{\bibfnamefont{J.}~\bibnamefont{Martin~Camalich}},
  \bibinfo{journal}{Phys. Rev. Lett.} \textbf{\bibinfo{volume}{118}},
  \bibinfo{pages}{081802} (\bibinfo{year}{2017}), \eprint{1611.06676}.

\bibitem[{\citenamefont{Akeroyd and Chen}(2017)}]{Akeroyd:2017mhr}
\bibinfo{author}{\bibfnamefont{A.~G.} \bibnamefont{Akeroyd}} \bibnamefont{and}
  \bibinfo{author}{\bibfnamefont{C.-H.} \bibnamefont{Chen}}
  (\bibinfo{year}{2017}), \eprint{1708.04072}.

\bibitem[{\citenamefont{Hung et~al.}(1982)\citenamefont{Hung, Buras, and
  Bjorken}}]{Hung:1981pd}
\bibinfo{author}{\bibfnamefont{P.~Q.} \bibnamefont{Hung}},
  \bibinfo{author}{\bibfnamefont{A.~J.} \bibnamefont{Buras}}, \bibnamefont{and}
  \bibinfo{author}{\bibfnamefont{J.~D.} \bibnamefont{Bjorken}},
  \bibinfo{journal}{Phys. Rev.} \textbf{\bibinfo{volume}{D25}},
  \bibinfo{pages}{805} (\bibinfo{year}{1982}).

\bibitem[{\citenamefont{Valencia and Willenbrock}(1994)}]{Valencia:1994cj}
\bibinfo{author}{\bibfnamefont{G.}~\bibnamefont{Valencia}} \bibnamefont{and}
  \bibinfo{author}{\bibfnamefont{S.}~\bibnamefont{Willenbrock}},
  \bibinfo{journal}{Phys. Rev.} \textbf{\bibinfo{volume}{D50}},
  \bibinfo{pages}{6843} (\bibinfo{year}{1994}), \eprint{hep-ph/9409201}.

\bibitem[{\citenamefont{Pati and Salam}(1974)}]{Pati:1974yy}
\bibinfo{author}{\bibfnamefont{J.~C.} \bibnamefont{Pati}} \bibnamefont{and}
  \bibinfo{author}{\bibfnamefont{A.}~\bibnamefont{Salam}},
  \bibinfo{journal}{Phys. Rev.} \textbf{\bibinfo{volume}{D10}},
  \bibinfo{pages}{275} (\bibinfo{year}{1974}), \bibinfo{note}{[Erratum: Phys.
  Rev.D11,703(1975)]}.

\bibitem[{\citenamefont{Crivellin et~al.}(2018)\citenamefont{Crivellin, Greub,
  Saturnino, and Müller}}]{Crivellin:2018yvo}
\bibinfo{author}{\bibfnamefont{A.}~\bibnamefont{Crivellin}},
  \bibinfo{author}{\bibfnamefont{C.}~\bibnamefont{Greub}},
  \bibinfo{author}{\bibfnamefont{F.}~\bibnamefont{Saturnino}},
  \bibnamefont{and} \bibinfo{author}{\bibfnamefont{D.}~\bibnamefont{Müller}}
  (\bibinfo{year}{2018}), \eprint{1807.02068}.

\bibitem[{\citenamefont{Crivellin
  et~al.}(2017{\natexlab{b}})\citenamefont{Crivellin, Mueller, Signer, and
  Ulrich}}]{Crivellin:2017dsk}
\bibinfo{author}{\bibfnamefont{A.}~\bibnamefont{Crivellin}},
  \bibinfo{author}{\bibfnamefont{D.}~\bibnamefont{Mueller}},
  \bibinfo{author}{\bibfnamefont{A.}~\bibnamefont{Signer}}, \bibnamefont{and}
  \bibinfo{author}{\bibfnamefont{Y.}~\bibnamefont{Ulrich}}
  (\bibinfo{year}{2017}{\natexlab{b}}), \eprint{1706.08511}.

\bibitem[{\citenamefont{Aaboud et~al.}(2017)}]{Aaboud:2017buh}
\bibinfo{author}{\bibfnamefont{M.}~\bibnamefont{Aaboud}} \bibnamefont{et~al.}
  (\bibinfo{collaboration}{ATLAS}), \bibinfo{journal}{JHEP}
  \textbf{\bibinfo{volume}{10}}, \bibinfo{pages}{182} (\bibinfo{year}{2017}),
  \eprint{1707.02424}.

\bibitem[{\citenamefont{Crivellin
  et~al.}(2016{\natexlab{b}})\citenamefont{Crivellin, D'Ambrosio, Hoferichter,
  and Tunstall}}]{Crivellin:2016vjc}
\bibinfo{author}{\bibfnamefont{A.}~\bibnamefont{Crivellin}},
  \bibinfo{author}{\bibfnamefont{G.}~\bibnamefont{D'Ambrosio}},
  \bibinfo{author}{\bibfnamefont{M.}~\bibnamefont{Hoferichter}},
  \bibnamefont{and} \bibinfo{author}{\bibfnamefont{L.~C.}
  \bibnamefont{Tunstall}}, \bibinfo{journal}{Phys. Rev.}
  \textbf{\bibinfo{volume}{D93}}, \bibinfo{pages}{074038}
  (\bibinfo{year}{2016}{\natexlab{b}}), \eprint{1601.00970}.

\bibitem[{\citenamefont{Fajfer et~al.}(2012)\citenamefont{Fajfer, Kamenik, and
  Nisandzic}}]{Fajfer:2012vx}
\bibinfo{author}{\bibfnamefont{S.}~\bibnamefont{Fajfer}},
  \bibinfo{author}{\bibfnamefont{J.~F.} \bibnamefont{Kamenik}},
  \bibnamefont{and}
  \bibinfo{author}{\bibfnamefont{I.}~\bibnamefont{Nisandzic}},
  \bibinfo{journal}{Phys.Rev.} \textbf{\bibinfo{volume}{D85}},
  \bibinfo{pages}{094025} (\bibinfo{year}{2012}), \eprint{1203.2654}.

\bibitem[{\citenamefont{Na et~al.}(2015)\citenamefont{Na, Bouchard, Lepage,
  Monahan, and Shigemitsu}}]{Na:2015kha}
\bibinfo{author}{\bibfnamefont{H.}~\bibnamefont{Na}},
  \bibinfo{author}{\bibfnamefont{C.~M.} \bibnamefont{Bouchard}},
  \bibinfo{author}{\bibfnamefont{G.~P.} \bibnamefont{Lepage}},
  \bibinfo{author}{\bibfnamefont{C.}~\bibnamefont{Monahan}}, \bibnamefont{and}
  \bibinfo{author}{\bibfnamefont{J.}~\bibnamefont{Shigemitsu}}
  (\bibinfo{collaboration}{HPQCD}), \bibinfo{journal}{Phys. Rev.}
  \textbf{\bibinfo{volume}{D92}}, \bibinfo{pages}{054510}
  (\bibinfo{year}{2015}), \bibinfo{note}{[Erratum: Phys.
  Rev.D93,no.11,119906(2016)]}, \eprint{1505.03925}.

\bibitem[{\citenamefont{Crivellin et~al.}(2015)\citenamefont{Crivellin, Hofer,
  Matias, Nierste, Pokorski et~al.}}]{Crivellin:2015era}
\bibinfo{author}{\bibfnamefont{A.}~\bibnamefont{Crivellin}},
  \bibinfo{author}{\bibfnamefont{L.}~\bibnamefont{Hofer}},
  \bibinfo{author}{\bibfnamefont{J.}~\bibnamefont{Matias}},
  \bibinfo{author}{\bibfnamefont{U.}~\bibnamefont{Nierste}},
  \bibinfo{author}{\bibfnamefont{S.}~\bibnamefont{Pokorski}},
  \bibnamefont{et~al.} (\bibinfo{year}{2015}), \eprint{1504.07928}.

\bibitem[{\citenamefont{Lees et~al.}(2012)}]{Lees:2012zz}
\bibinfo{author}{\bibfnamefont{J.~P.} \bibnamefont{Lees}} \bibnamefont{et~al.}
  (\bibinfo{collaboration}{BaBar}), \bibinfo{journal}{Phys. Rev.}
  \textbf{\bibinfo{volume}{D86}}, \bibinfo{pages}{012004}
  (\bibinfo{year}{2012}), \eprint{1204.2852}.

\bibitem[{\citenamefont{Bobeth et~al.}(2014)\citenamefont{Bobeth, Gorbahn,
  Hermann, Misiak, Stamou, and Steinhauser}}]{Bobeth:2013uxa}
\bibinfo{author}{\bibfnamefont{C.}~\bibnamefont{Bobeth}},
  \bibinfo{author}{\bibfnamefont{M.}~\bibnamefont{Gorbahn}},
  \bibinfo{author}{\bibfnamefont{T.}~\bibnamefont{Hermann}},
  \bibinfo{author}{\bibfnamefont{M.}~\bibnamefont{Misiak}},
  \bibinfo{author}{\bibfnamefont{E.}~\bibnamefont{Stamou}}, \bibnamefont{and}
  \bibinfo{author}{\bibfnamefont{M.}~\bibnamefont{Steinhauser}},
  \bibinfo{journal}{Phys. Rev. Lett.} \textbf{\bibinfo{volume}{112}},
  \bibinfo{pages}{101801} (\bibinfo{year}{2014}), \eprint{1311.0903}.

\bibitem[{\citenamefont{Bobeth}(2014)}]{Bobeth:2014tza}
\bibinfo{author}{\bibfnamefont{C.}~\bibnamefont{Bobeth}}, in
  \emph{\bibinfo{booktitle}{{Proceedings, 49th Rencontres de Moriond on
  Electroweak Interactions and Unified Theories: La Thuile, Italy, March 15-22,
  2014}}} (\bibinfo{year}{2014}), pp. \bibinfo{pages}{75--80},
  \eprint{1405.4907},
  \urlprefix\url{http://inspirehep.net/record/1297237/files/arXiv:1405.4907.pdf}.

\bibitem[{\citenamefont{Bobeth et~al.}(2000)\citenamefont{Bobeth, Misiak, and
  Urban}}]{Bobeth:1999mk}
\bibinfo{author}{\bibfnamefont{C.}~\bibnamefont{Bobeth}},
  \bibinfo{author}{\bibfnamefont{M.}~\bibnamefont{Misiak}}, \bibnamefont{and}
  \bibinfo{author}{\bibfnamefont{J.}~\bibnamefont{Urban}},
  \bibinfo{journal}{Nucl. Phys.} \textbf{\bibinfo{volume}{B574}},
  \bibinfo{pages}{291} (\bibinfo{year}{2000}), \eprint{hep-ph/9910220}.

\bibitem[{\citenamefont{Huber et~al.}(2006)\citenamefont{Huber, Lunghi, Misiak,
  and Wyler}}]{Huber:2005ig}
\bibinfo{author}{\bibfnamefont{T.}~\bibnamefont{Huber}},
  \bibinfo{author}{\bibfnamefont{E.}~\bibnamefont{Lunghi}},
  \bibinfo{author}{\bibfnamefont{M.}~\bibnamefont{Misiak}}, \bibnamefont{and}
  \bibinfo{author}{\bibfnamefont{D.}~\bibnamefont{Wyler}},
  \bibinfo{journal}{Nucl. Phys.} \textbf{\bibinfo{volume}{B740}},
  \bibinfo{pages}{105} (\bibinfo{year}{2006}), \eprint{hep-ph/0512066}.

\bibitem[{\citenamefont{Aaij et~al.}(2017{\natexlab{b}})}]{Aaij:2017xqt}
\bibinfo{author}{\bibfnamefont{R.}~\bibnamefont{Aaij}} \bibnamefont{et~al.}
  (\bibinfo{collaboration}{LHCb}) (\bibinfo{year}{2017}{\natexlab{b}}),
  \eprint{1703.02508}.

\bibitem[{\citenamefont{Alonso et~al.}(2015)\citenamefont{Alonso, Grinstein,
  and Camalich}}]{Alonso:2015sja}
\bibinfo{author}{\bibfnamefont{R.}~\bibnamefont{Alonso}},
  \bibinfo{author}{\bibfnamefont{B.}~\bibnamefont{Grinstein}},
  \bibnamefont{and} \bibinfo{author}{\bibfnamefont{J.~M.}
  \bibnamefont{Camalich}} (\bibinfo{year}{2015}), \eprint{1505.05164}.

\bibitem[{\citenamefont{Choudhury et~al.}(2017)\citenamefont{Choudhury, Kundu,
  Mandal, and Sinha}}]{Choudhury:2017qyt}
\bibinfo{author}{\bibfnamefont{D.}~\bibnamefont{Choudhury}},
  \bibinfo{author}{\bibfnamefont{A.}~\bibnamefont{Kundu}},
  \bibinfo{author}{\bibfnamefont{R.}~\bibnamefont{Mandal}}, \bibnamefont{and}
  \bibinfo{author}{\bibfnamefont{R.}~\bibnamefont{Sinha}}
  (\bibinfo{year}{2017}), \eprint{1706.08437}.

\bibitem[{\citenamefont{Falkowski et~al.}(2014)\citenamefont{Falkowski, Straub,
  and Vicente}}]{Falkowski:2013jya}
\bibinfo{author}{\bibfnamefont{A.}~\bibnamefont{Falkowski}},
  \bibinfo{author}{\bibfnamefont{D.~M.} \bibnamefont{Straub}},
  \bibnamefont{and} \bibinfo{author}{\bibfnamefont{A.}~\bibnamefont{Vicente}},
  \bibinfo{journal}{JHEP} \textbf{\bibinfo{volume}{05}}, \bibinfo{pages}{092}
  (\bibinfo{year}{2014}), \eprint{1312.5329}.

\bibitem[{\citenamefont{Dermisek et~al.}(2014)\citenamefont{Dermisek, Hall,
  Lunghi, and Shin}}]{Dermisek:2014qca}
\bibinfo{author}{\bibfnamefont{R.}~\bibnamefont{Dermisek}},
  \bibinfo{author}{\bibfnamefont{J.~P.} \bibnamefont{Hall}},
  \bibinfo{author}{\bibfnamefont{E.}~\bibnamefont{Lunghi}}, \bibnamefont{and}
  \bibinfo{author}{\bibfnamefont{S.}~\bibnamefont{Shin}},
  \bibinfo{journal}{JHEP} \textbf{\bibinfo{volume}{12}}, \bibinfo{pages}{013}
  (\bibinfo{year}{2014}), \eprint{1408.3123}.

\bibitem[{\citenamefont{Belanger et~al.}(2015)\citenamefont{Belanger, Delaunay,
  and Westhoff}}]{Belanger:2015nma}
\bibinfo{author}{\bibfnamefont{G.}~\bibnamefont{Belanger}},
  \bibinfo{author}{\bibfnamefont{C.}~\bibnamefont{Delaunay}}, \bibnamefont{and}
  \bibinfo{author}{\bibfnamefont{S.}~\bibnamefont{Westhoff}},
  \bibinfo{journal}{Phys. Rev.} \textbf{\bibinfo{volume}{D92}},
  \bibinfo{pages}{055021} (\bibinfo{year}{2015}), \eprint{1507.06660}.

\bibitem[{\citenamefont{Altmannshofer et~al.}(2016)\citenamefont{Altmannshofer,
  Carena, and Crivellin}}]{Altmannshofer:2016oaq}
\bibinfo{author}{\bibfnamefont{W.}~\bibnamefont{Altmannshofer}},
  \bibinfo{author}{\bibfnamefont{M.}~\bibnamefont{Carena}}, \bibnamefont{and}
  \bibinfo{author}{\bibfnamefont{A.}~\bibnamefont{Crivellin}},
  \bibinfo{journal}{Phys. Rev. Lett.}  (\bibinfo{year}{2016}),
  \bibinfo{note}{[Phys. Rev.D94,095026(2016)]}, \eprint{1604.08221}.

\bibitem[{\citenamefont{Chen et~al.}(2016)\citenamefont{Chen, Nomura, and
  Okada}}]{Chen:2016dip}
\bibinfo{author}{\bibfnamefont{C.-H.} \bibnamefont{Chen}},
  \bibinfo{author}{\bibfnamefont{T.}~\bibnamefont{Nomura}}, \bibnamefont{and}
  \bibinfo{author}{\bibfnamefont{H.}~\bibnamefont{Okada}},
  \bibinfo{journal}{Phys. Rev.} \textbf{\bibinfo{volume}{D94}},
  \bibinfo{pages}{115005} (\bibinfo{year}{2016}), \eprint{1607.04857}.

\bibitem[{\citenamefont{Arnan et~al.}(2016)\citenamefont{Arnan, Hofer, Mescia,
  and Crivellin}}]{Arnan:2016cpy}
\bibinfo{author}{\bibfnamefont{P.}~\bibnamefont{Arnan}},
  \bibinfo{author}{\bibfnamefont{L.}~\bibnamefont{Hofer}},
  \bibinfo{author}{\bibfnamefont{F.}~\bibnamefont{Mescia}}, \bibnamefont{and}
  \bibinfo{author}{\bibfnamefont{A.}~\bibnamefont{Crivellin}}
  (\bibinfo{year}{2016}), \eprint{1608.07832}.

\bibitem[{\citenamefont{Coluccio~Leskow
  et~al.}(2016)\citenamefont{Coluccio~Leskow, Crivellin, D'Ambrosio, and
  Mueller}}]{ColuccioLeskow:2016dox}
\bibinfo{author}{\bibfnamefont{E.}~\bibnamefont{Coluccio~Leskow}},
  \bibinfo{author}{\bibfnamefont{A.}~\bibnamefont{Crivellin}},
  \bibinfo{author}{\bibfnamefont{G.}~\bibnamefont{D'Ambrosio}},
  \bibnamefont{and} \bibinfo{author}{\bibfnamefont{D.}~\bibnamefont{Mueller}}
  (\bibinfo{year}{2016}), \eprint{1612.06858}.

\bibitem[{\citenamefont{Di~Chiara et~al.}(2017)\citenamefont{Di~Chiara, Fowlie,
  Fraser, Marzo, Marzola, Raidal, and Spethmann}}]{DiChiara:2017cjq}
\bibinfo{author}{\bibfnamefont{S.}~\bibnamefont{Di~Chiara}},
  \bibinfo{author}{\bibfnamefont{A.}~\bibnamefont{Fowlie}},
  \bibinfo{author}{\bibfnamefont{S.}~\bibnamefont{Fraser}},
  \bibinfo{author}{\bibfnamefont{C.}~\bibnamefont{Marzo}},
  \bibinfo{author}{\bibfnamefont{L.}~\bibnamefont{Marzola}},
  \bibinfo{author}{\bibfnamefont{M.}~\bibnamefont{Raidal}}, \bibnamefont{and}
  \bibinfo{author}{\bibfnamefont{C.}~\bibnamefont{Spethmann}}
  (\bibinfo{year}{2017}), \eprint{1704.06200}.

\bibitem[{\citenamefont{Di~Luzio et~al.}(2017)\citenamefont{Di~Luzio, Greljo,
  and Nardecchia}}]{DiLuzio:2017vat}
\bibinfo{author}{\bibfnamefont{L.}~\bibnamefont{Di~Luzio}},
  \bibinfo{author}{\bibfnamefont{A.}~\bibnamefont{Greljo}}, \bibnamefont{and}
  \bibinfo{author}{\bibfnamefont{M.}~\bibnamefont{Nardecchia}}
  (\bibinfo{year}{2017}), \eprint{1708.08450}.

\bibitem[{\citenamefont{Assad et~al.}(2017)\citenamefont{Assad, Fornal, and
  Grinstein}}]{Assad:2017iib}
\bibinfo{author}{\bibfnamefont{N.}~\bibnamefont{Assad}},
  \bibinfo{author}{\bibfnamefont{B.}~\bibnamefont{Fornal}}, \bibnamefont{and}
  \bibinfo{author}{\bibfnamefont{B.}~\bibnamefont{Grinstein}}
  (\bibinfo{year}{2017}), \eprint{1708.06350}.

\end{thebibliography}

\end{document}